\documentclass{article}

\usepackage{arxiv}

\usepackage[utf8]{inputenc} 
\usepackage[T1]{fontenc}    
\usepackage{hyperref}       
\usepackage{url}            
\usepackage{booktabs}       
\usepackage{amsfonts}       
\usepackage{nicefrac}       
\usepackage{microtype}      
\usepackage{lipsum}		
\usepackage{graphicx}
\usepackage{doi}

\usepackage{comment}

\usepackage{amsmath}

\usepackage{amsmath}

\title{Machine learning memory kernels as closure for non-Markovian stochastic processes}

\author{Antonio~Russo \\
Department of Chemical Engineering\\
Imperial College London\\
London SW7 2AZ, UK\\
\And
Miguel~A.~Dur\'{a}n-Olivencia \\
Department of Chemical Engineering\\
Imperial College London\\
London SW7 2AZ, UK \\
\And
Ioannis G. Kevrekidis \\
Department of Chemical and Biomolecular Engineering\\
Johns Hopkins University\\
Baltimore MD 21218, USA\\
\And
Serafim~Kalliadasis \\
Department of Chemical Engineering\\
Imperial College London\\
London SW7 2AZ, UK \\
}

\date{}

\renewcommand{\headeright}{}
\renewcommand{\undertitle}{}

\renewcommand{\shorttitle}{Machine learning memory kernels as closure for non-Markovian stochastic processes}

\begin{document}
	
	\maketitle
	


%




\begin{abstract}
Finding the dynamical law of observable quantities lies at the core of
physics.
Within the particular field of statistical mechanics, the generalized
Langevin equation (GLE) comprises a general model for the evolution of
observables covering a great deal of physical systems with many degrees of
freedom and an inherently stochastic nature.
Although formally exact, the GLE brings its own great challenges.
It depends on the complete history of the observables under scrutiny, as
well as the microscopic degrees of freedom, all of which are often
inaccessible.
We show that these drawbacks can be overcome by adopting elements of
machine learning from empirical data, in particular coupling a multilayer
perceptron (MLP) with the formal structure of the GLE and calibrating the
MLP with the data.
This yields a powerful computational tool capable of describing noisy
complex systems beyond the realms of statistical mechanics.
It is exemplified with a number of representative examples from different
fields: from a single colloidal particle and particle chains in a thermal
bath to climatology and finance, showing in all cases excellent agreement
with the actual observable dynamics.
The new framework offers an alternative perspective for the study of
non-equilibrium processes opening also a new route for stochastic
modelling.
\end{abstract}


%

\section{Introduction}

\begin{figure*} [!t]
	\centering
	\includegraphics[width=0.9\textwidth]{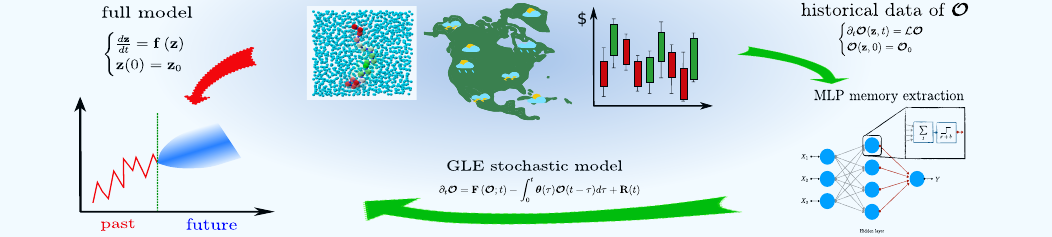}
	\caption{Two possible approaches to simulate the time evolution of dynamical-systems' observable.
		As a first approach, the full deterministic dynamical system (red arrow) is solved.
		Despite the advantage of being exact, this approach is often not suitable either because is computationally prohibitive
		or because a model of the full DoF system is not accessible.
		An alternative consists of building a GLE model for the observable and parameterizing
		it through a multilayer perceptron, given appropriate historical data (green arrows).
		Suitable system observables can be selected based either on prior knowledge of the system or directly on historical data by means of
		techniques such as principal-component analysis, auto-encoders and diffusion maps\cite{Singer2009,Ferguson2010}.
	}
	\label{fig:main}	
\end{figure*}

The mathematical description of both natural and technological processes
requires governing equations for the observed temporal evolution of pertinent
process properties.
Such quantities can often be directly measured, and act as a bridge between theory and experiments.
Consequently, they are known as \emph{observables}.
Physical systems can be analyzed following either a microscopic description, i.e. a complete description of each component in the system, or a macroscopic approach, i.e studying the system as a whole.
	The set of observables, $\boldsymbol{\mathcal{O}}=\{\mathcal{O}_i\}_{_{i=1,2,\dots}}$, which
	uniquely describes the macroscopic state of a system is typically termed as
	\textit{canonical observables}, e.g. pressure and temperature to describe a
	thermodynamic state.
	At the same time, the minimum set of variables, $\mathbf{z}=\{z_j\}_{_{j=1,2,\dots}}$, required to describe the microscopic state of a dynamical system is referred
	to as \textit{degrees of freedom} (DoF).
	Statistical mechanics deals with the connection between macroscopic
	observables (or simply observables) and DoF.
From a purely theoretical point of view, any observable can be understood as
a function of the system's DoF, i.e. $\mathcal{O}_i =f_i(\mathbf{z}) $.
Given the huge number of DoF a physical system typically involves, finding
the exact functional form $f$ which connects the given observable and the
system's DoF represents an overwhelming challenge.
It is formally possible to go from the DoF description of the system (e.g. from
Newton's equations of motion) to one in terms of observables via
dimensionality reduction which retains the main effects at the observables'
level and allows us to describe the same phenomenon with a substantially
reduced number of variables.
Such a reduction has the convenience of a simpler representation of
the system, enabling also its study in a computationally inexpensive manner.
This is of central importance for the understanding of complex systems, given
the huge computational cost required to integrate the DoF over time in
realistic scenarios.
Indeed, the number of DoF is typically as large as
Avogadro's number ($N_{\text{A}}\sim 10^{23}$).
%
Unfortunately, postulating a dynamical law for an observable is in general non-trivial.
This necessitates the quest for finding the connection between the DoF
and the observable dynamics.

There exists a mathematical formalism which allows us to get the formal
structure of the equations describing the observable dynamics starting from
the deterministic DoF's time-evolution equations without knowing exactly the
functional dependency of the observables on the DoF.
This dimensionality-reduction formalism is known as the ``projection-operator
(PO) technique", and was originally introduced by Mori and
Zwanzig\cite{Mori1965,Zwanzig1973} and comprehensive analysis of the
formalism in Ref.\cite{Chorin2006} where reduction techniques are reviewed.
Despite not yielding closed governing equations (precisely because of the
inherent limitation of not knowing the functional relationship $f$ between
the observables and the DoF), it produces a relatively simple and versatile
model after some convenient simplifications.
The first success of the PO formalism was the derivation of the Brownian
dynamics, for which only a phenomenological derivation by Langevin was
available at the time\cite{Lemons1997}.

\subsection{Mori-Zwanzig's formalism}
	Let us consider the following (linear of non-linear) deterministic dynamical system:
	\begin{align}
	\begin{cases}
	\frac{d  \mathbf{z}}{d t} = \mathbf{f} \left( \mathbf{z} \right) \\
	\mathbf{z}(0)  = \mathbf{z}_0
	\end{cases}
	\label{GLE:eq:dynamical_system1}
	\end{align}
	where $\mathbf{z} \in \mathbb{R}^n$ is a vector of independent variables. 
	For the system in Eq.~\eqref{GLE:eq:dynamical_system1}, it can be defined a set of observables $\boldsymbol{ \mathcal{O} }(\mathbf{z},t)= \phi(\mathbf{z}(t))$, where $\phi$ represent the transformation map between $\mathbf{z}$ and $\boldsymbol{ \mathcal{O} }$. 
	By using the chain rule, it is easy to show that the evolution equation of $\boldsymbol{ \mathcal{O} }(\mathbf{z},t)$ can be written as:
	\begin{align}
	\begin{cases}
	\frac{\partial  \boldsymbol{ \mathcal{O} } }{\partial t} ( \mathbf{z} , t )= \mathcal{L} \boldsymbol{ \mathcal{O} } \\
	\boldsymbol{ \mathcal{O} } ( \mathbf{z} , 0 )  = \boldsymbol{ \mathcal{O} }_0
	\end{cases}
	\label{GLE:eq:observable_dynamics1}
	\end{align}
	where it was introduced the operator $ \mathcal{L} = \mathbf{f} \left( \mathbf{z} \right) \cdot \nabla_{\mathbf{z}}$.
	It follows that the solution of Eq.~\eqref{GLE:eq:observable_dynamics1} can be written as: 
	\begin{align}
	\boldsymbol{ \mathcal{O} } ( \mathbf{z} , t ) = e^{\mathcal{L} t} \boldsymbol{ \mathcal{O} }_0 
	\label{GLE:eq:observable_dynamics_sol}
	\end{align}
	where the exponential has to be intended as the power series that defines the exponential map between matrices.

If we are only interested in the dynamics of some observables $\boldsymbol{ \mathcal{O} }$, rather then the whole solution $\mathbf{z}(t)$, we can define a projection operator $\mathcal{P}$, which maps functions of $\mathbf{z}$ into function of $\boldsymbol{ \mathcal{O} }$. 
	It is worth underlining that, in general, the set of observables $\boldsymbol{ \mathcal{O} }$ may be defined by a linear or nonlinear transformation of $\mathbf{z}$, but in any case the evolution of $\boldsymbol{ \mathcal{O} }$ is supposed to be unitary, i.e. $\mid \boldsymbol{ \mathcal{O} }(t) \mid^2 = \mid \boldsymbol{ \mathcal{O} }(0) \mid^2 $. 
	A simple, but still important, scenario is given by $\boldsymbol{ \mathcal{O} }$ being a subset of $\mathbf{z}$. 
	As we will see later, this case plays a fundamental role in dimensional reductions of multi-component systems, i.e. colloidal particles in a thermal bath.
	Given a projection operator $\mathcal{P}$, namely a transformation from a vector space to itself such that $\mathcal{P}^2=\mathcal{P}$, one can follow Mori-Zwanzig's formalism~\cite{Zwanzig1960,Mori1965,Zwanzig1973} to obtain a form of Eq.~\eqref{GLE:eq:observable_dynamics1} suitable for system dimensionality reduction. 
	Note that at this point no constrain is put on the form of the projection operator.
	After defining the operator $\mathcal{Q} = \mathbf{1} - \mathcal{P}$, orthogonal to $\mathcal{P}$, Eq.~\eqref{GLE:eq:observable_dynamics1} can be rewritten as:
	\begin{align}
	\frac{\partial  \boldsymbol{ \mathcal{O} } }{\partial t} ( \mathbf{z} , t ) = \mathcal{L} e^{\mathcal{L} t} \boldsymbol{ \mathcal{O} }_0 = e^{\mathcal{L} t} \mathcal{P} \mathcal{L} \boldsymbol{ \mathcal{O} }_0 + e^{\mathcal{L} t} \mathcal{Q} \mathcal{L} \boldsymbol{ \mathcal{O} }_0 
	\label{GLE:eq:observable_dynamics11}
	\end{align}
	Duhamel-Dyson's formula allows to rewrite the exponential term $e^{\mathcal{L} t}$ as:
	\begin{align}
	e^{\mathcal{L} t} = e^{\mathcal{Q} t} + \int_0^t e^{\mathcal{L}(t-\tau)} \ \mathcal{P} \  e^{\mathcal{Q}\tau} \ d \tau
	\label{GLE:eq:dyson}
	\end{align}
	and, consequently, Eq.~\eqref{GLE:eq:observable_dynamics11} becomes the so called Mori-Zwanzig's equation:
	\begin{align}
	\begin{split}
	\frac{\partial  \boldsymbol{ \mathcal{O} } }{\partial t} ( \mathbf{z} , t ) =  & e^{\mathcal{L} t} \mathcal{P} \mathcal{L} \boldsymbol{ \mathcal{O} }_0 +  \\
	&\int_0^t e^{\mathcal{L}(t-\tau)}  \mathcal{P} \mathcal{L} e^{\mathcal{Q}\tau}  \mathcal{Q} \mathcal{L} \boldsymbol{ \mathcal{O} }_0 \ d \tau + e^{\mathcal{Q} \mathcal{L}  t} \mathcal{Q} \mathcal{L} \boldsymbol{ \mathcal{O} }_0 
	\label{GLE:eq:observable_dynamics2}
	\end{split}
	\end{align}
	The first term is the Markovian contribution, the second constitutes the memory term and the last one is often interpreted as the noise. 
	It is worth noticing that, at this stage, Eq.~\eqref{GLE:eq:observable_dynamics2} is exactly equivalent to Eq.~\eqref{GLE:eq:dynamical_system1} and is valid independently from the specific choice of the projection operator $\mathcal{P}$.
	Mori and Zwanzig~\cite{Mori1965,Zwanzig1973,Zwanzig2001} proposed two different projection operators leading to different forms of GLE, that we will briefly discuss in next sections. 

If we name the noise term $ \mathbf{R}(t) =  e^{\mathcal{Q} \mathcal{L} t} \mathcal{Q} \mathcal{L} \boldsymbol{ \mathcal{O} }_0$, then the following dynamical system remains determined:
	\begin{align}
	\begin{cases}
	\dfrac{\partial  \mathbf{R} }{\partial t}  ( \boldsymbol{ \mathcal{O} }_0 , t ) = \mathcal{Q} \mathcal{L} \mathbf{R} ( \boldsymbol{ \mathcal{O} }_0 , t ) , \\
	\mathbf{R} ( \boldsymbol{ \mathcal{O} }_0 , t ) = \mathcal{Q} \mathcal{L} \boldsymbol{ \mathcal{O} }_0 .
	\end{cases}
	\label{GLE:eq:noise_term}
	\end{align}
	Projecting Eq.~\eqref{GLE:eq:noise_term} according to $\mathcal{P}$, it follows:
	\begin{align}
	\begin{cases}
	\mathcal{P} \dfrac{\partial  \mathbf{R} }{\partial t}  ( \boldsymbol{ \mathcal{O} }_0 , t ) = \mathcal{P} \mathcal{Q} \mathcal{L} \mathbf{R} ( \boldsymbol{ \mathcal{O} }_0 , t ) = \mathbf{0} , \\
	\mathcal{P} \mathbf{R} ( \boldsymbol{ \mathcal{O} }_0 , t ) = \mathcal{P} \mathcal{Q} \mathcal{L} \boldsymbol{ \mathcal{O} }_0 = \mathbf{0} ,
	\end{cases}
	\label{GLE:eq:projected_noise_term}
	\end{align}
	where we have used the property of the projection operator $\mathcal{P}\mathcal{Q}=\mathbf{0}$. 
	This shows that $\mathbf{R}(t)$ is orthogonal to the range of $\mathcal{P}$ at any time $t$. 
	However, in order to express $\mathbf{R}(t)$ as a stochastic process, it is necessary to have either time scale separation or weak coupling between resolved and unresolved variables~\cite{Givon2004}.
	When at least one of such conditions occurs, at least asymptotically, the influence of the unresolved variables may be interpreted as sum of many uncorrelated events, and consequently can be treated with Central Limit Theorem~\cite{Gottwald2017}. 
	Thus, it is the Central Limit Theorem that determines the Gaussian shape for the distribution of $\mathbf{R}(t)$, while its time correlation follows from the fluctuation dissipation theorem, as shown in what follows. 
	\paragraph{Mori's projection operator}
	The projection operator introduced by Mori~\cite{Mori1965}, when applied to a general variable $\mathbf{A} (\mathbf{z})$, is defined as:
	\begin{align}
	\mathcal{P} \mathbf{A} (\mathbf{z}) = \langle \mathbf{A} , \boldsymbol{ \mathcal{O} }_0 \rangle \langle  \boldsymbol{ \mathcal{O} }_0, \boldsymbol{ \mathcal{O} }_0 \rangle^{-1} \boldsymbol{ \mathcal{O} }_0
	\label{GLE:eq:mori_operator}
	\end{align}
	where the inner product $\langle \mathbf{A} , \mathbf{B} \rangle$ is defined as
	\begin{align}
	\langle \mathbf{A} , \mathbf{B}  \rangle = \int \rho (\mathbf{z}) \mathbf{A} (\mathbf{z})  \mathbf{B}^* (\mathbf{z}) \ d \mathbf{z}
	\end{align}
	with $\rho(\mathbf{z}) $ being a normalized probability density function defined in the phase space of the original system and $\mathbf{B}^*$ the conjugate transpose of $\mathbf{B}$. 
	In case of systems with Hamiltonian $\mathcal{H}$ in a canonical ensemble, the probability density function is $\rho(\mathbf{z})= Z^{-1} e^{-\beta \mathcal{H}(\mathbf{z})} $, where $Z$ is the partition function and $\beta= k_B T $.
	Employing Mori's operator in Eq.~\eqref{GLE:eq:observable_dynamics2}, we obtain the Markovian term:
	\begin{align}
	e^{\mathcal{L} t} \mathcal{P} \mathcal{L} \boldsymbol{ \mathcal{O} }_0 = \langle \mathcal{L} \boldsymbol{ \mathcal{O} }_0 , \boldsymbol{ \mathcal{O} }_0 \rangle \langle \boldsymbol{ \mathcal{O} }_0, \boldsymbol{ \mathcal{O} }_0 \rangle^{-1} \boldsymbol{ \mathcal{O} } (t) .
	\end{align}
	Moreover, from the definition of $\mathbf{R}(t)$, we obtain the memory term:
	\begin{align}
	\int_0^t e^{\mathcal{L}(t-\tau)}  \mathcal{P} \mathcal{L} e^{\mathcal{Q}\tau}  \mathcal{Q} \mathcal{L} \boldsymbol{ \mathcal{O} }_0 \ d \tau = - \int_0^t \boldsymbol{\theta}(\tau) \boldsymbol{ \mathcal{O} } (t-\tau) \ d \tau
	\end{align}
	where the memory kernel is defined as $\boldsymbol{\theta}(t)= - \langle \mathcal{L} \mathbf{R}(t) , \boldsymbol{ \mathcal{O} }_0 \rangle \langle \boldsymbol{ \mathcal{O} }_0, \boldsymbol{ \mathcal{O} }_0 \rangle^{-1}$. 
	Since $\mathcal{Q} \mathbf{R}(t)= \mathbf{R}(t)$, and $\mathcal{L}$ is an anti-Hermitian operator~\cite{Zwanzig2001}, it follows that $\langle \mathcal{L} \mathbf{R}(t) , \boldsymbol{ \mathcal{O} }_0 \rangle= - \langle \mathbf{R}(t) , \mathcal{L} \boldsymbol{ \mathcal{O} }_0 \rangle= - \langle  \mathbf{R}(t) , \mathcal{Q} \mathcal{L} \boldsymbol{ \mathcal{O} }_0 \rangle = - \langle  \mathbf{R}(t) , \mathbf{R}(0) \rangle $.
	Hence, we obtain the following relation:
	\begin{align}
	\boldsymbol{\theta}(t)= \langle  \mathbf{R}(t) ,  \mathbf{R}(0) \rangle \langle \boldsymbol{ \mathcal{O} }_0, \boldsymbol{ \mathcal{O} }_0 \rangle^{-1} ,
	\end{align}
	which constitutes the fluctuation dissipation theorem. 
	\paragraph{Zwanzig projection operator}
	As Zwanzig pointed out, Mori's projection operator leads to a linearised GLE~\cite{Zwanzig2001}. 
	Zwanzig~\cite{Zwanzig1960,Zwanzig2001} defined the projection operator applied to the variable $\mathbf{A} (\mathbf{z})$ through the following conditional expectation:
	\begin{align}
	\mathcal{P} \mathbf{A} (\mathbf{z}) = \dfrac{ \int \rho(\mathbf{z}) \ \mathbf{A}(\mathbf{z}) \ \delta ( \boldsymbol{ \mathcal{O} } - \phi(\mathbf{z}) ) \ d \mathbf{z}} {  \int \rho(\mathbf{z}) \ \delta ( \boldsymbol{ \mathcal{O} } - \phi(\mathbf{z})) \ d \mathbf{z}} ,
	\label{GLE:eq:zwanzig_operator}
	\end{align}
	where $ \delta ( \boldsymbol{ \mathcal{O} } - \phi(\mathbf{z})) = \prod_j \delta ( \boldsymbol{ \mathcal{O} }_j - \phi_j)$. 
	%
	In molecular dynamics, the set of observables is often defined as a subset of the original coordinates, namely $\boldsymbol{ \mathcal{O} } \subseteq \mathbf{z}$.
	In this case, Zwanzig's projection operator allows to express the Markovian term in Eq.~\eqref{GLE:eq:observable_dynamics2} as function of the potential of mean force. 
	To show this, let us consider an isothermal Hamiltonian system of $N$ particles with coordinates $\mathbf{z}=\{\mathbf{r},\mathbf{p} \}$, where $\mathbf{r}= \{ \mathbf{r}_1 \dots \mathbf{r}_N \} $ and $\mathbf{p} = \mathbf{p}_1 \dots \mathbf{p}_N $ are position and momenta, respectively. 
	With $\mathbf{f} \left( \mathbf{z} \right)=  -\mathbf{\nabla}_{\mathbf{z}} V(\mathbf{z})$, Eq.~\eqref{GLE:eq:dynamical_system1} gives the Newton's equations of motion for a system of interacting particles. 
	Suppose one is interested in the dynamical evolution of only $n$ of the original $N$ particles, whose coordinates (called relevant variables) and are indicated as $ \tilde{\mathbf{z} } = \{ \mathbf{r}_1 \dots \mathbf{r}_n , \mathbf{p}_1 \dots \mathbf{p}_n \}$. 
	The remaining variables, called unresolved variables, are denoted by $ \hat{\mathbf{z} } = \{ \mathbf{r}_{n+1} \dots \mathbf{r}_N , \mathbf{p}_{n+1} \dots \mathbf{p}_N \}$.
	Hence, inserting Zwanzig's operator in Eq.~\eqref{GLE:eq:observable_dynamics2}, we obtain the Markovian term in the form:
	\begin{align}
	\mathcal{P} \mathcal{L} \tilde{\mathbf{z} }   = \dfrac{ \int \ -\mathbf{\nabla}_{\mathbf{z}} V(\mathbf{z})  e^{-\beta \mathcal{H (\mathbf{z}) }} \ \delta ( \mathbf{z} - \tilde{\mathbf{z} } ) \ d \mathbf{z} } {  \int e^{-\beta \mathcal{H (\mathbf{z}) } } \ \delta ( \mathbf{z} - \tilde{\mathbf{z} } ) \ d \mathbf{z} } = -\mathbf{\nabla}_{\mathbf{\tilde{\mathbf{z} } }} V^{\text{PMF}} (\tilde{\mathbf{z} } )
	\label{GLE:potential_mean_force}
	\end{align}
	where $V^{\text{PMF}}$ is known as potential of mean force.
	Moreover, the memory term can be written in terms of the noise term as:
	\begin{align}
	\int_0^t e^{\mathcal{L}(t-\tau)}  \mathcal{P} \mathcal{L} \mathbf{R} (\tau) \ d \tau .
	\label{GLE:eq:memory_term}
	\end{align}
	Ref.~\cite{Chen2014} has shown that the term in Eq.~\eqref{GLE:eq:memory_term} is null for the position coordinates $\mathbf{r}$, while can expressed for the momentum coordinates $\mathbf{p}$ as the convolution $- \int_{0}^{t}  \boldsymbol{\theta}(\tau) \mathbf{p} (t-\tau) d \tau $.

\paragraph{Generalized Langevin Equation}
In short, the PO formalism makes it possible to derive the
time-evolution equation for $\boldsymbol{\mathcal{O}}$,
which results in a set of first-order generalized Langevin equations (GLEs).
The structure of a GLE typically consists of a: 
\begin{itemize}
	\item Markovian (mean-force) term which depends on the instantaneous
	values of the observables under consideration~\cite{Kinjo2007,Hijon2010,Chen2014}, 
	\item Non-Markovian (so-called memory) time-convolution term which depends on the
	historical values of the observables at hand. In many cases the Non-Markovian term can be expressed simply as the convolution between
	the observables and a tensor function (so-called memory kernel), $\boldsymbol{\theta} (t)$, which plays the role of a viscosity
	kernel~\cite{Kinjo2007,Hijon2010,Chen2014}. 
	\item Noise term which depends on both the observables and all DoF's
	initial conditions, and which is why it is interpreted as a purely
	random term.
\end{itemize}
%

The PO formalism makes it possible to assert that any set of observables will
follow the GLE dynamics:
\begin{align}
\partial_t \boldsymbol{ \mathcal{O} }(t) = \mathbf{F}(\boldsymbol{ \mathcal{O} } (t) )- \int_{0}^{t}  \boldsymbol{\theta}(\tau) \boldsymbol{ \mathcal{O} } (t-\tau) d \tau + \mathbf{R}(t) .
\label{eq:gle}
\end{align}
with $\mathbf{F}( \boldsymbol{ \mathcal{O} } (t))$ being the deterministic term, and $\mathbf{R}(t)$ the stochastic
term orthogonal to $\boldsymbol{ \mathcal{O} }$, with correlation given by the
fluctuation-dissipation theorem:
\begin{equation}
\langle \mathbf{R}(t) , \mathbf{R}(t') \rangle =
\boldsymbol{\theta}(t-t') \langle \boldsymbol{\mathcal{O}} , \boldsymbol{\mathcal{O}} \rangle.
\label{eq:flu-dis-rel}
\end{equation}
In Eq. \eqref{eq:flu-dis-rel}
we have used $\langle a , b\rangle = \int d\mathbf{z}\, \rho(\mathbf{z})
a(\mathbf{z}) b^\dag(\mathbf{z})$ where $\mathbf{z} \in \mathbb{R}^n$ is the
system's DoF and $\rho(\mathbf{z})$ a normalized probability density
function, and $b^\dag$ the conjugate transpose of $b$.
%
%
As can be readily checked, the non-Markovian term depends on the temporal
trace of the system and is characterised by the memory-kernel function
$\boldsymbol{\theta}(t)$.
This function also determines the noise term $\mathbf{R}(t)$ through the
fluctuation-dissipation theorem given above.
Therefore, ascertaining the memory kernel is crucial for preserving the main
features of the high-dimensional (microscopic) dynamics into the
dimensionally-reduced time-evolution equations.
Unfortunately, the memory kernel depends on the whole set of DoF and their
full history\cite{Grabert1982} which makes the problem intractable as
illustrated in Fig.~\ref{fig:main}.
Because of these difficulties, more often than not, the memory kernel
$\boldsymbol{\theta}(t)$ is approximated through the Markovian hypothesis,
$\boldsymbol{\theta}(t)= \boldsymbol{\theta}_0 \delta(t)$.
However, this approximation as much as it yields drastically simpler LEs, it
comes at the expense of accuracy representing considerable source of errors
as we will demonstrate.

\subsection{Main highlights and applications}

Analytical expressions for the memory kernel are only accessible for very
specific, and often ideal, systems such as the academic case of a particle in
a harmonic-oscillator heat bath~\cite{Zwanzig2001}. But in the general case
it cannot be determined from first principles and analysis can only take us
so far. A good example is given in Ref.~\cite{Chorin2000} where the
authors adopt a perturbation scheme which is yet ``too complex for general
use". Numerical techniques are the only way out when dealing with realistic
systems, e.g. systems where non-linear interactions prevail.
To obtain information on what is proven to be an unfathomable quantity, a
handful of recent studies have focused on numerical parametrization
techniques aiming to decipher its main features. In principle, this can
enable the numerical simulation of stochastic systems with a low
computational cost compared to other methods [e.g. traditional agent-based
simulations such as molecular dynamics (MD)].

Alas, the numerical front is not free of challenges. Despite its accuracy,
the algorithm developed in Ref.~\cite{Darve2009} to parameterize GLEs
involves sampling of the full high-dimensional system.
Such an algorithm is $O(N)$-complex, which is not ideal for practical
purposes as the computational costs grows (at least linearly) with the number
of particles in the system.
In Ref.~\cite{Torres-Carbajal2015}, an iterative
approach is adopted to compute a discrete/point-wise approximation of
$\boldsymbol{\theta}(t)$ from the system's autocorrelation functions.
However, this makes the convolution and the stochastic term of the GLEs
computationally intractable, as they both depend on $\boldsymbol{\theta}(t)$.
Finally, in Refs~\cite{Lange2006,Lei2016} the authors propose to
extract the memory kernel by Laplace transforming the correlation functions
computed from historical data of the observables.
While certainly interesting, this strategy exhibits serious limitations when
the available data of the observables' dynamics are affected by even small
fluctuations as we shall demonstrate.

Our overarching objective here is the development of a novel data-driven
approach where the memory kernel is machine learned from observation data.
For this purpose we propose the use of a feed-forward artificial neural
network, namely a multilayer perceptron (MLP), to achieve an efficient and
systematic forecasting of the memory kernel.
We provide the MLP with appropriate historical data of the observables under
study, obtained either from simulations or public databases.
The MLP is then trained via an optimization process to approximate the memory
kernel with a degree of accuracy depending on the number of neurons
in the hidden layer.
In particular, the memory kernel is extracted as an expansion of
multi-exponential functions, which allows us to derive a tractable stochastic
integration algorithm of a non-Markovian process characterized by
time-correlated noise.

In this work, we present a novel data-driven approach, which makes use of the GLE structure coupled with a multilayer perceptron (MLP) to achieve an optimal parametrization of the memory kernel. 
	The MLP is provided with proper historical data of the observables of interest obtained either from simulations or existing databases and then executes an optimization procedure to find the optimal approximation of the memory kernel. 
	As we shown later in this section, compared to previous approaches our approximation through MLP shows enhanced robustness, especially when the available data are limited or affected by significant fluctuations.
	In the this procedure, the memory kernel is extracted in the form of a multi-exponential functions, thus enabling us to derive a tractable stochastic integration algorithm of the non-Markovian process characterized by a time-correlated noise. 
	The universal approximation theorem~\cite{Hornik1989,Hornik1991} guarantees a wide applicability of the our methodology which is tested in some relevant case studies from chemistry, biology, climatology and finance. 
	These include: Modelling the dynamics of a single colloidal particle immersed in a heat bath of identical particles, also in presence of an external potential; Coarse-graining a particle chain in a bath; Modelling historical trends of a financial index.

\label{method}

Consider a system which has reached statistical equilibrium characterized by
a stationary distribution $\rho(\mathbf{z})$.
The historical data of the observables of the system,
$\boldsymbol{\mathcal{O}}$, can be then be considered as a realization of a
stationary process.
With the aim of obtaining a direct relationship between
$\boldsymbol{\theta}(t)$ and the statistics of the evolution of
$\boldsymbol{\mathcal{O}}$, we can take the inner product of
Eq.~\eqref{eq:gle} and the observables' initial condition $\boldsymbol{
	\mathcal{O} }(0)$, yielding:
\begin{align}
\mathbf{g}(t) =& - \left\langle \partial_t \boldsymbol{ \mathcal{O} }(t) - \mathbf{F}(\boldsymbol{ \mathcal{O} } (t) ), \boldsymbol{ \mathcal{O} }(0) \right\rangle  \nonumber\\
= & \left\langle \int_{0}^{t}  \boldsymbol{\theta}(\tau) \boldsymbol{ \mathcal{O} } (t-\tau) d \tau + \mathbf{R}(t), \boldsymbol{ \mathcal{O} }(0) \right\rangle
\nonumber\\
=& \int_{0}^{t} \boldsymbol{\theta} (t-\tau) \mathbf{h} (\tau) d \tau ,
\label{eq:gle_correlation}
\end{align}
where
	$\mathbf{h}(t)= \langle \boldsymbol{ \mathcal{O} }(t) , \boldsymbol{\mathcal{O} }(0) \rangle$, $\mathbf{g}(t)- \left\langle \partial_t \boldsymbol{ \mathcal{O} }(t) - \mathbf{F}(\boldsymbol{ \mathcal{O} } (t) ), \boldsymbol{ \mathcal{O} }(0) \right\rangle  \nonumber$ is implicitly defined by the relation above, and $\langle
	\mathbf{R}(t),\boldsymbol{ \mathcal{O} }(0) \rangle \sim \mathbf{0}$ because of the orthogonality between the random force and the initial value of the observables.
For some specific cases (e.g. one-dimensional systems, or colloidal systems
of spherical particles), $\boldsymbol{\theta}(t)= \theta(t) \mathbf{1}$,
$\mathbf{g}(t)= g(t) \mathbf{1}$ and $\mathbf{h}(t)= h(t) \mathbf{1}$.
For the sake of convenience we will then drop the identity operator
$\mathbf{1}$ in such cases and simply use $\theta(t)$, $g(t)$ and $h(t)$.

\subsection{Memory kernel in the Laplace space}

\begin{figure}[t]
	\centering
	\begin{minipage}{0.4\textwidth}
		\flushleft (a)\\
		\centering	
		\includegraphics[width=\textwidth]{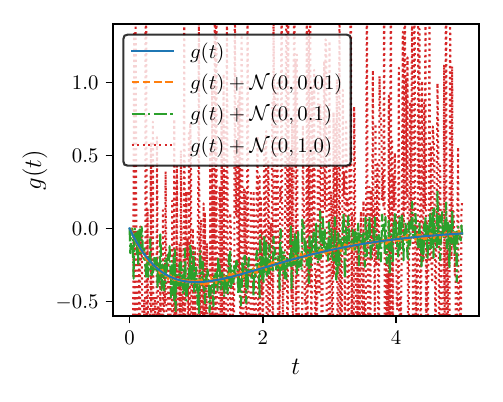}
	\end{minipage}
	\begin{minipage}{0.4\textwidth}
		\flushleft (b)\\
		\centering	
		\includegraphics[width=\textwidth]{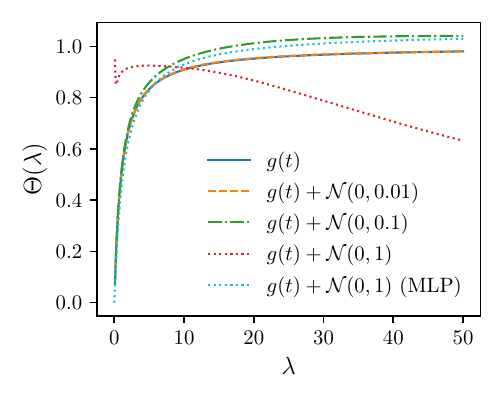}
	\end{minipage}
	\begin{minipage}{0.4\textwidth}
		\flushleft (c)\\
		\centering	
		\includegraphics[width=\textwidth]{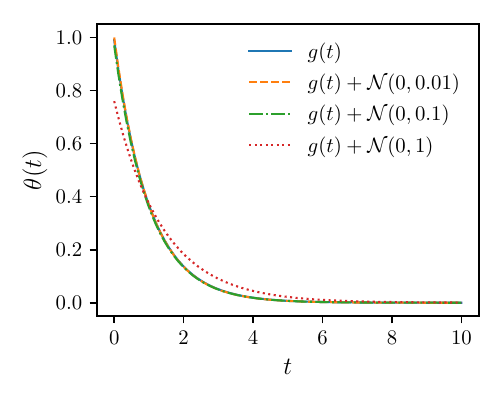}
	\end{minipage}
	\caption{Convolution function $g(t)$ affected by random noise with varying amplitudes (a).
		Comparison between the memory kernel $\theta$ computed in the Laplace space (b) and with our MLP-based method (c),
		for $g(t)$ affected by a random noise with several amplitudes.
		For comparison purposes, we also report in (b) the Laplace transform of the memory kernel obtained with our MLP for strong noise,
		i.e. for noise variance $\sigma^2=1$.
	}
	\label{fig:memory_kernel_noise}
\end{figure}

Equation~\eqref{eq:gle_correlation} can be rewritten by utilizing the properties of Laplace transform ($\mathcal{L} \left\{\boldsymbol{\mu}\right\}(\lambda)= \int_0^{\infty} \boldsymbol{\mu}(t) e^{-t / \lambda}$) which turns the convolution integral into a
multiplication:
\begin{equation}
\mathbf{G}(\lambda) =
\boldsymbol{\Theta}(\lambda)\,\mathbf{H}(\lambda) \Rightarrow
\boldsymbol{\Theta}(\lambda)  =  \mathbf{G}(\lambda)\,\mathbf{H}(\lambda)^{-1}
\end{equation}
where $\mathbf{G}(\lambda),\ \boldsymbol{\Theta}(\lambda)$ and
$\mathbf{H}(\lambda)$ are the Laplace transforms of $\mathbf{g}(t)$,
$\boldsymbol{\theta}(t)$ and $\mathbf{h}(t)$, respectively.
One approach is to adopt a rational function approximation for $\boldsymbol{\Theta}(\lambda)$ to be fitted to data, as was done by Lei \emph{et al}.\cite{Lei2016}
This approximation seems to behave well in the absence of noise in the data.
However, unfortunately this approach fails for limited data,
which produce correlations affected by random noise.
It is straightforward to demonstrate this serious limitation. Consider a function $\mathbf{g}(t)$
affected by a Gaussian source of uncertainty, $\boldsymbol{\epsilon}(t)$.
This leads to an error in the Laplace transform, which can be expressed as
$\Delta \boldsymbol{\Theta}(\lambda) = \tilde{\boldsymbol{\Theta}}(\lambda) -
\boldsymbol{\Theta}(\lambda) =
\boldsymbol{\mathcal{E}}(\lambda)\,\mathbf{H}^{-1}(\lambda)$, where
$\boldsymbol{\mathcal{E}}=\mathcal{L}\left\{\boldsymbol{\epsilon}(t)\right\}$.
For the sake of argument, assume that the Gaussian uncertainty can be
expressed as the sum of non-systematic local errors, i.e.
$\boldsymbol{\epsilon}(t) = \sum_i \boldsymbol{\epsilon}_i \delta (t-t_i)$,
with $ \boldsymbol{\epsilon}_i \sim \mathcal{N}( 0, \sigma^2)$.
Therefore, $\Delta \boldsymbol{\Theta} (\lambda)  =  \sum_i
\boldsymbol{\epsilon}_i e^{-t_i / \lambda} \mathbf{H}^{-1}(\lambda)$.
This clearly shows that local temporal errors turn into non-local
contributions in the Laplace space.
Such an \emph{error propagation} will inevitably lead to significant
inaccuracies that will compromise the rational approximation of the memory
kernel.
For instance, take the simple case $h(t)= e^{-t}$ which straightforwardly
enables the analytical treatment of Eq.~\eqref{eq:gle_correlation}, $g(t)= -t
e^{-t}$ (see Fig.~\ref{fig:memory_kernel_noise}(a)).
Figure~\ref{fig:memory_kernel_noise}(b) shows that the Laplace transform of
the memory kernel (as computed in Ref.~\cite{Lei2016}) diverges from
the actual curve as the noise intensity increases.
Indeed the Laplace transform should be avoided.
We shall demonstrate that the adoption of an MLP-based procedure provides an
efficient and robust approximation of $\boldsymbol{\theta}$ in the time
domain. Indeed, as illustrated in Fig.~\ref{fig:memory_kernel_noise}(c), the
expected memory kernel is approximated very well even in the presence of
strong noise in the underlying data.

\subsection{Memory kernel extraction through MLP}

Amongst the different possible neural network structures, MLPs have gained
popularity because of their versatility and capability in approximating
highly-nonlinear functions\cite{Nielsen2015}.
An MLP consists of at least three layers (known as input, hidden and output
layers), each of them including several nodes or neurons
(Fig.~\ref{fig:main}).
The transformation of the dataset at each node is determined by an activation function, $\phi$.
Every node's inputs are weighted and added together with a bias, effectively an offset, to be
finally passed through the activation function to compute its output.
Training the network consists of finding the optimal weights and the biases
which lead to minimize an appropriate cost (or error) function, computed at
the output of the MLP.
A popular choice for the cost function is the mean-square error (according to
either Euclidean or non-Euclidean metrics).
The iterations of the optimization process are known as epochs, and the whole
procedure to find the optimal MLP parameterization is referred to as the
\emph{learning process}.
We propose a three-layer MLP as a typical architecture for the
multidimensional nonlinear regression problem of estimating the memory kernel.
It should be highlighted that standard approaches model directly system data series (e.g. Ref.~\cite{Hochreiter1997}), while our approach is different in that it integrates our physical knowledge of the phenomena at hand to extend the range of applicability of the model.
As an example, consider a system in a known bistable potential for which we have access to data on system configurations in a single well only.
With a standard approach, a machine learning model trained with these data would fail to detect bistability or estimate the transition time distribution.
On the contrary, in our approach, a GLE embedded with a bi-stable potential and the extracted memory-kernel would be able to accurately model bi-stability and transition dynamics of such system.

The input of the MLP is a vector containing discrete time values, while the
desired output is the function $g(t)$, which is known \textit{a priori}.
The hidden layer has an arbitrary number of neurons, determining the degree
of accuracy of the memory-kernel approximation.
As activation functions we use $\phi(z)=\int_{0}^{t} h (t-\tau) e^{z(\tau)} d
\tau$ in the hidden layer, with $h(t)$ being known a priori and $\phi(z)=z$
at the output layer.
The learning algorithm adopted is the resilient back-propagation algorithm with an adaptive learning rate.\cite{Riedmiller1994}
Providing the MLP with the two matrices $\mathbf{g}(t)$ and $\mathbf{h}(t)$,
an optimal approximator is obtained upon completion of the learning process.
Such an estimator can be expressed as an exponential series, namely as $
\boldsymbol{\theta}(t) \sim \sum_{k=1}^{N_n} \mathbf{A}_k e^{ \mathbf{B}_k
	t} $, where $N_n$ is the number of nodes in the hidden layer, $
\mathbf{A}_k$ are matrices of real numbers and $ \mathbf{B}_k$ are matrices
with real negative coefficients (more details are given in the Supplementary Information).
We now comment on the choice of the particular MLP configuration. This is
because a three-layer MLP is the minimal network architecture satisfying the
conditions of the universal approximation theorem.
According to such a theorem, this network structure is able to approximate
any continuous function defined on a compact subset of
$\mathbb{R}^d$.\cite{Hornik1989,Hornik1991}
Unfortunately, there are no formal results on the number of nodes required in
the hidden layer to ensure the proper learning of the memory kernel function.
But systematic experimentation gives the number that best worked in our case studies.
It might seem counterintuitive, given the complexity of the prototypical
examples we will consider, or even misleading, that the number of nodes needed
in our experiments is just a few neurons.
The main reason for this is that the major computational challenge would have
been the learning of the dynamical law.
However, this step is already made by the use of statistical mechanics which
establishes the GLE as a general equation for the evolution of observable
quantities and the description on non-Markovian Gaussian processes, making it
possible to focus the computational effort on the learning of only a
particular ingredient of the whole problem, namely the memory kernel.
The simplest memory kernel one could employ is a Dirac delta function, which leads to the standard Langevin equation, i.e. the Markovian approximation of GLE.
A realistic way of modelling memory kernels is by an exponentially decaying function, which weights each historical configuration based on its distance in time from the current state of the system.
Approximating the memory kernel as a constant function is not realistic for physical systems as memory kernels represent the temporarily effects of a previous state of the system on the (implicit) environment.
It is then not surprising that our MLP suggests approximating the memory kernel by the sum of few exponentially decaying functions. This works with a high degree of accuracy for complex systems.

\subsection{Multi-layer perceptron structure and learning algorithm}
\label{MLP}

Artificial neural networks are used for the parameterization of the GLE
because of their enhanced capabilities to model non-linear relationships
between system variables. Developed by analogy with biological processes in
the brain, artificial neural networks are series of linear and non-linear
transformations of some inputs to some outputs.
Amongst the different possible variants, multi-layer perceptrons (MLPs) have
gained popularity because of their potential and versatility in non-linear
function approximations\cite{Nielsen2015,Hochreiter1997}.
MLPs consist of at least three layers, known as input, hidden and output
layers, each one including more nodes.
Each node $i$ in the layer $l-1$ is connected with any other node $j$ in the
successive layer $l$ and every connection is characterized by a parameter
$w_{j,i}^l$ known as weight.
In addition, for every neuron in the network there is a parameter known as
the bias, $b_j^l$.
The transformation of the dataset at each node is determined by an activation
function, $\phi(z_j^l )$.
It follows that the output $a_j^l$ of the neuron $j$ of the layer $l$ is
computed as $a_j^l= \phi \left( z_j^l \right)$, with $z_j^l=\sum_{i}
w_{j,i}^l a_i^{l-1} +b_j \label{eq:a_j^l}$.
The network learning process then consists of an optimization algorithm
aiming to find the weights $w_{j,i}^l$ and the biases $b_j^l$ that minimize a
cost (or error) function $C$ computed at the output of the MLP.
In this work, we employ a quadratic cost function $C= \sum_{t_j}^{N_t} \frac{1}{2 N_t} \left( y_{j}(t_j)- a_j^{L}(t_j) \right)^2$ ,
where $N_t$ is the number of data samples.
Hence, an algorithm is used to cyclically back-propagate the information about the error evaluated at the output to update weights and bias.
%

We adopt a three-layer MLP with a single input and a single output function.
The hidden layer has an arbitrary number of neurons, $N_n$, determining the
degree of accuracy of the memory kernel approximation.
The universal approximation theorem guarantees that such a structure of the
network is able to approximate any continuous function defined on a compact
subset of $\mathbb{R}^d$\cite{Hornik1989,Hornik1991}.
Initialization of the MLPs is achieved by providing Gaussian distributed
random numbers to the weights and zeros of the biases.
Moreover, no bias is added at the output layer.
Regarding the activation function, in the hidden layer we adopt
$\phi(z)=\int_{0}^{t} h (t-\tau) e^{z (\tau)} d \tau$, with $h(t)$ being
known a priori, while at the output layer we employ $\phi(z)= z$.

For the learning process we adopt the resilient back-propagation algorithm
Rprop\cite{Riedmiller1994} based on the gradient descent method, ${\mathbf
	{\alpha}}_{e+1}={\mathbf {\alpha}}_e -\eta \nabla_{\mathbf{\alpha}}
C({\mathbf{\alpha}}_e ) $,
with the adaptive learning rate $\eta$,
\begin{align}
\eta (e)=\begin{cases}
\eta^+ \cdot \eta(e-1) & \text{if} \quad \frac{\partial C}{\partial \alpha}(e) \cdot \frac{\partial C}{\partial \alpha}(e-1) > 0 ,\\
\eta^- \cdot \eta (e-1) & \text{if} \quad \frac{\partial C}{\partial \alpha}(e) \cdot \frac{\partial C}{\partial \alpha}(e-1) < 0 ,\\
\eta (e-1) & \text{otherwise} ,
\end{cases}
\label{eq:Rprop}
\end{align}
where $\mathbf{\alpha}=\left[ w_{j,i}^l ; b_{j}^l \right]$, and $0< \eta^-<1<\eta^+$
are fixed parameters.
From experience and following the literature\cite{Riedmiller1994}, the Rprop
algorithm gives an optimal compromise between calculation speed and solution
convergence.
Providing the MLP with the two matrices $\mathbf{g(t)}$ and $\mathbf{h(t)}$,
the memory kernel is then extracted in the form of an exponential series:
\begin{align}
\theta(t) \sim  \sum_{k=1}^{N_n} w_k^{3} e^{b_k^{2}} e^{w_k^{2} t} = \sum_{k=1}^{N_n} A_k e^{B_k (t)} ,
\end{align}
where $N_n$ is the number of nodes in the hidden layer, $A_k=w_i^{3}
e^{b_k^{2}}$ are real coefficients and $B_k=w_k^{2}$ are real strictly
negative quantities.
The algorithm presented so far has been adopted to extract the memory kernel
in the case of a diagonal $\boldsymbol{\theta}(t)$.
%

\subsection{GLE time integration}
\label{GLE_integration}
The integration of the GLE dynamics is a non-trivial task for two reasons.
First, the convolution integral depends on the full history of the
observables. And second, the stochastic term is correlated in time.
Different approaches have been proposed to address these issues for the scalar case \cite{Baczewski2013,Kawai2015,Lei2016}.
In this work, we take
advantage of the exponential structure of the identified
$\boldsymbol{\theta}(t)$ to implement an integration algorithm. The
history-dependent convolution term is then written as a sum of the additional
variables $ \mathbf{Z}_{k}(t)$, each defined as $ \mathbf{Z}_{k}(t)=
\int_{0}^{t} \mathbf{A}_k e^{ \mathbf{B}_k (t-\tau)} \boldsymbol{ \mathcal{O}
} (\tau) d \tau $, so that their evolution equation can be expressed as
$\dot{ \mathbf{Z}_{k}}(t)= \mathbf{B}_k \mathbf{Z}_{k}(t) - \mathbf{A}_k
\boldsymbol{ \mathcal{O} }(t)$. The noise $ \mathbf{R}(t)$ must satisfy the
fluctuation-dissipation theorem. By introducing a set of auxiliary
variables $ \boldsymbol{\xi}_k (t)$, we can rewrite $ \mathbf{R}(t)= \sum_{k=1}^{Nn} \mathbf{R}_k(t) =
\sum_{k=1}^{Nn} \mathbf{b}_k \boldsymbol{\xi}(t)$, so that the corresponding
evolution reads $\dot{ \mathbf{R}_{k}}(t)= \mathbf{B}_k \mathbf{R}_{k}(t) +
\mathbf{b}_k \boldsymbol{\xi}(t) $, where $ \boldsymbol{\xi} (t)$ is a white
noise with zero mean and time correlation $\langle \boldsymbol{\xi} (t)
\boldsymbol{\xi} (s) \rangle = 2 \langle \boldsymbol{ \mathcal{O} },
\boldsymbol{ \mathcal{O} }\rangle \ \delta(t-s)$, while the coefficients $
\mathbf{b}_k$ can be computed numerically (for details see Supplementary Information).
By defining the variables $ \mathbf{S}_k(t)= - \mathbf{Z}_k(t) +
\mathbf{R}_k(t)$, the GLE can then  be rewritten in the extended form:
\begin{align}
\begin{cases}
\dot{\boldsymbol{ \mathcal{O} }} (t) &= \mathbf{F}(\boldsymbol{ \mathcal{O} }(t)) + \sum_{k=1}^{N_n} \mathbf{S}_{k}(t) \\
\dot{ \mathbf{S}_{k}}(t) &= \mathbf{B}_k \mathbf{S}_{k}(t) - \mathbf{A}_k \boldsymbol{ \mathcal{O} } (t) + \mathbf{b}_k \boldsymbol{\xi} (t) ,
\end{cases}
\label{extended_dynamics}
\end{align}
with $ \mathbf{F} (\boldsymbol{ \mathcal{O} }(t))$ accounting for the mean force contributions.

\section{Numerical applications}

\subsection{Single particle in a bath}

\begin{figure*}[!h]
	\centering
	\begin{minipage}{0.4\textwidth}
		\flushleft (a)\\
		\centering	
		\includegraphics[width=\textwidth]{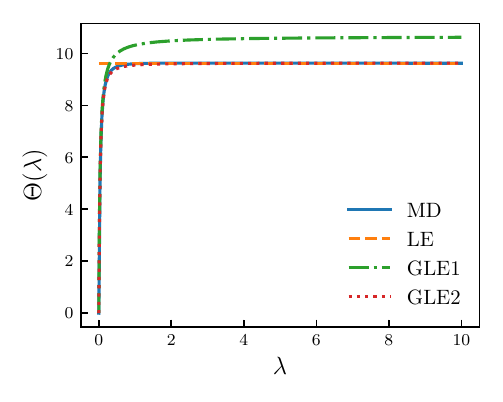}
	\end{minipage}
	\begin{minipage}{0.4\textwidth}
		\flushleft (b)\\
		\centering	
		\includegraphics[width=\textwidth]{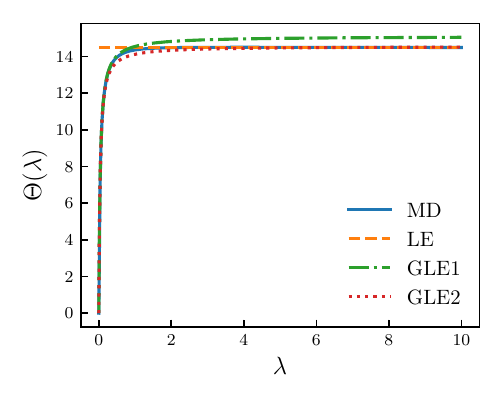}
	\end{minipage}
	\begin{minipage}{0.4\textwidth}
		\flushleft (c)\\
		\centering
		\includegraphics[width=\textwidth]{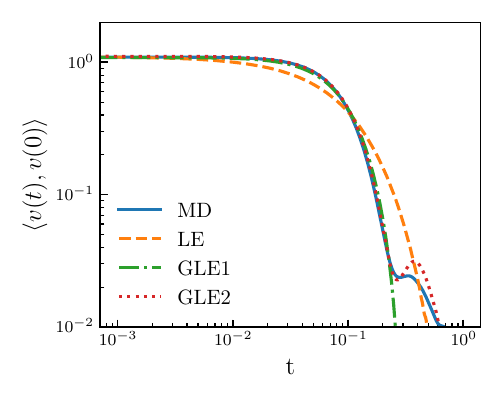}
	\end{minipage}
	\begin{minipage}{0.4\textwidth}
		\flushleft (d)\\
		\centering
		\includegraphics[width=\textwidth]{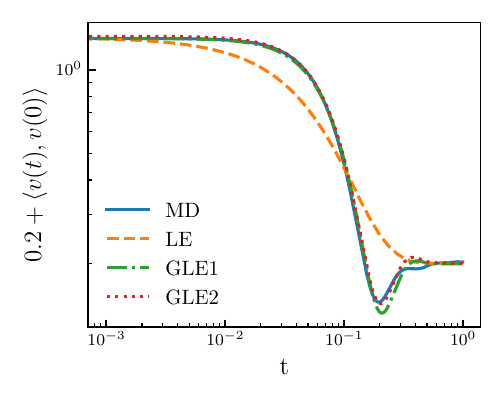}
	\end{minipage}
	\begin{minipage}{0.4\textwidth}
		\flushleft (e)\\	
		\centering	
		\includegraphics[width=\textwidth]{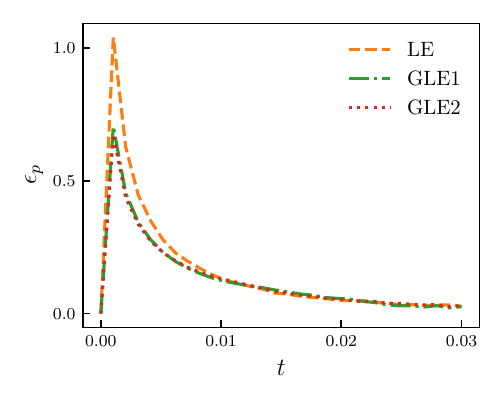}
	\end{minipage}
	\begin{minipage}{0.4\textwidth}
		\flushleft (f)\\	
		\centering	
		\includegraphics[width=\textwidth]{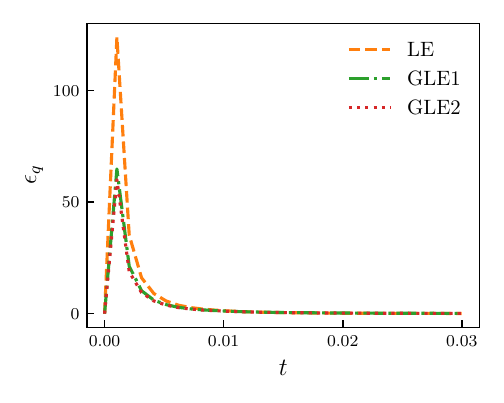}
	\end{minipage}
	\caption{Memory kernels approximated through the MLP are compared against the ones obtained directly from MD simulation in the Laplace space, for (a) LDL and (b) HDL cases.
		Velocity correlation functions computed from MD, Langevin equation (LE) and GLE dynamics over $10^4$ trajectories for both LDL (c) and HDL (d) cases are also reported.
		GLE1 and GLE2 refer to the memory kernel approximations obtained, respectively, with 1 and 2 neurons in the hidden layer.
		In (c) a constant factor of $0.2$ is added to avoid negative values of the correlation in the log-log plot.
		%
		In (e-f) we report the mean square differences $\epsilon_p(t)$ and $\epsilon_q(t)$ between the PDFs of the reduced systems (GLE and LE) and the exact pdf of the full system (MD) as function of the relaxation time.
	}
	\label{fig:bath_theta}
\end{figure*}

The first test to exemplify our methodology is a well-studied problem: the
dynamics of a single colloidal particle (with mass $m_B$) immersed in a heat
bath of $n_b$ identical particles (with mass $m_b$).
This problem also serves classically as model prototype for the derivation of
the LE.
The observable to be modelled with the MLP-enriched GLE of Eq.\eqref{eq:gle}
is the mean velocity of the colloidal particle, $v(t)$.
while the historical data to be used for the training of the MLP are the
values of the momenta of the target particle and forces acting on it
generated with equilibrium molecular dynamics (MD) simulations.
The interaction between two particles $i$ and $j$ is modelled by the
Lennard-Jones (LJ) potential:
\begin{align}
v_{\text{LJ}}(r)=
\begin{cases}
4 \epsilon  \left[ \left( \sigma / r\right)^{12} - \left( \sigma / r \right)^{6}  \right] & \ \text{if} \ r \leq r_c , \\
0 &\ \text{otherwise},
\end{cases}
\label{eq:lennard_jones}
\end{align}
where $r$ is the distance between the particles, $\epsilon$ is the depth of
the potential well, $\sigma$ is the finite distance at which the
inter-particle potential is zero and $r_c=2.5\,\sigma$ is a
cut-off radius.
The numerical results are reported in reduced units, using $\sigma$ and
$\epsilon$ to scale lengths, energies and times, respectively.
Our MD set-up is a cubic box of length $L=10\,\sigma$ (hence volume
$V=10^3$), periodic boundary conditions along the Cartesian coordinates, $x,
y$ and $z$, and a Nos\'e-Hoover thermostat to equilibrate the system at a
reduced temperature $T=1.0$ (equivalent to $k_BT = \epsilon$).
We consider two different scenarios depending on the bath particle densities:
the low density limit (LDL) with $\rho_b=\frac{n_b}{V}=\frac{699}{1000}$,
and the high density limit (HDL) with $\rho_b=\frac{799}{1000}$.
The comparisons of the MLP-estimated and the exact/MD-extracted
Laplace-transform of the memory kernels are shown in
Figs~\ref{fig:bath_theta}(a)-(b) under LDL and HDL conditions, respectively.
The use of the Laplace transform here is merely for comparison purposes,
since we actually extract the memory kernel $\boldsymbol{\theta} (t)$ from
our MD data.
As can be readily checked in Figs~\ref{fig:bath_theta}(a)-(b), the
first-order MLP approximator (GLE1) obtained with a single neuron at the
hidden layer already outperforms the Markovian approximation (LE).
Despite being already quite good, this first-order approximation is still
unable to capture the behavior of $\boldsymbol{\Theta}(\lambda)$ for large
values of $\lambda$.
But just by adding a second neuron at the hidden layer, the second-order
approximator (GLE2) perfectly converges to the exact MD results over the the
whole $\lambda$-axis.
In Table \ref{table:convergence_single_particle} we report the cost function error value after training.
It can be noted that increasing the number of nodes in the hidden layer above $2$ does not increase the accuracy of the MLPs approximation and additional nodes are redundant.
As expected, the accuracy of the approximations (LE, GLE1 and GLE2) has a
direct impact on the velocity correlation $\langle v(t), v(0)\rangle$
obtained, as shown in Figs~\ref{fig:bath_theta}(c)-(d).
These figures clearly demonstrate the limitations of the Markovian
approximation, which is quite different compared to the actual correlation decay.
The first-order approximation is again fairly accurate, but diverges for long times.
On the other hand, the second-degree approximation follows the exact
autocorrelation within a tolerance lower than $1\%$.
Having a very good estimation of the actual $\boldsymbol{\theta}(t)$, we can
now proceed with the simulation of the reduced dynamics, i.e. the simulation
of the MLP-enriched GLE, and compare with that obtained by MD
simulations.
For this purpose we simulate the GLE and MD dynamics out of equilibrium under
the LDL.
We analyze the time evolution towards equilibrium of the probability-density
function (PDF) for the momentum and position, $\varrho_p$ and $\varrho_q$,
respectively, and as initial condition we chose a Dirac's delta distribution.
In Figs~\ref{fig:bath_theta}(e)-(f), we show the standard errors for the
momentum and position PDFs of the target particle, $\epsilon_\mu =
||\varrho_\mu^{\nu}-\varrho_\mu^{\text{MD}}||$ with $\mu\in\{p, q\}$ and
$\nu\in\{\text{LE, GLE1, GLE2}\}$.
As can be clearly seen in the figures, our MLP-based GLE method dramatically
reduces both errors $\epsilon_p$ and $\epsilon_q$ when compared to the
Markovian approximation, up to a $50\%$ lesser than LE during the
non-equilibrium relaxation.

\begin{table}
	\centering
	\caption{Value of the MLP error function after training for the LDL case.}
	\begin{tabular}{| c | c |}
		\hline
		Hidden nodes & Value of error function  \\ \hline
		1 & $6 \times 10^{-2}$   \\ \hline
		2 & $1 \times 10^{-2}$   \\ \hline
		3 & $1 \times 10^{-2}$   \\ \hline
		4 & $1 \times 10^{-2}$   \\ \hline
	\end{tabular}
	\label{table:convergence_single_particle}
\end{table}

\subsection{Particle in a bistable potential}

\begin{figure}[!h]
	\centering
	\begin{minipage}{0.4\textwidth}
		\flushleft (a)\\
		\centering	
		\includegraphics[width=\textwidth]{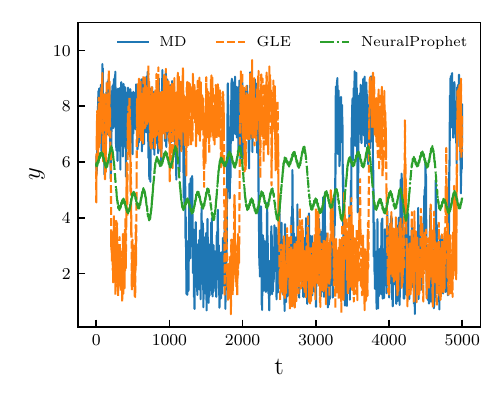}
	\end{minipage}
	\begin{minipage}{0.4\textwidth}
		\flushleft (b)\\
		\centering	
		\includegraphics[width=\textwidth]{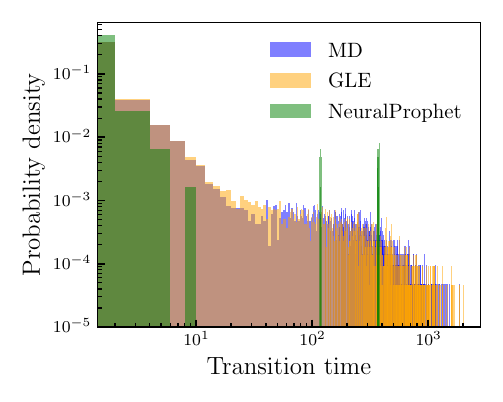}
	\end{minipage}
	\caption{(a) Trajectory of a particle in a bistable potential simulated with MD (explicit bath particles), GLE embedded with a memory kernel approximated through our MLP (implicit bath particles) and a standard package using neural network for data series modelling (NeuralProphet).
		The comparison between transition time probability densities (b) shows that NeuralProphet cannot accurately reproduce the transition dynamics of the particle because of the non-seasonal behaviour of the original MD data series.
		On the contrary, the transition time distribution obtained with our GLE follows very closely the MD one.
	}
	\label{fig:bath_bistable}
\end{figure}

As an additional validation of our GLE approach, we simulated a particle in a
bath confined in a double well potential $U^{ext} = 10 \times (y^4-y^2) $,
and compared the GLE dynamics against MD, and against a
neural-network forecasting of the dynamics by using  NeuralProphet (a standard
numerical library for data series modelling). NeuralProphet is based on an open-source
software used for time data series forecasts by Facebook's core data science team\cite{Taylor2017}.
It adopts an additive model where non-linear trends are fit with yearly, weekly,
and daily seasonality, plus holiday effects.
In Figure \ref{fig:bath_bistable}(a) we show a particle trajectory in the bistable potential simulated with MD (explicit bath particles), GLE embedded with a memory kernel approximated through our MLP (implicit bath particles) and NeuralProphet.
This visualization shows that NeuralProphet cannot accurately reproduce the transition dynamics of the particle because of the non-seasonal behaviour of the original (MD) data series.
In Figure \ref{fig:bath_bistable}(b), we also report the probability densities of the transition time, defined as the time difference between two consecutive crossings of the saddle point of the potential.
This comparison between MD, GLE and NeuralProphet confirms that standard packages,
such as NeuralProphet, cannot detect and replicate the full kinetics of transition
dynamics dominated by non-seasonal events.
On the contrary, our GLE approach shows its high capabilities in reproducing the MD transition time.

\subsection{Particle chain in a bath}

\begin{figure}[t!]
	\centering
	\begin{minipage}{0.4\textwidth}
		\centering
		\includegraphics[width=\textwidth]{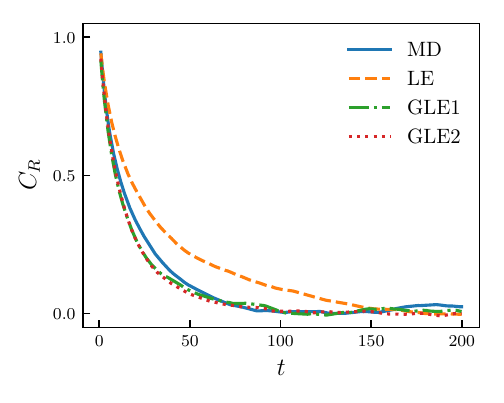}
	\end{minipage}
	\caption{Time correlation of the gyration radius of a particle chain in a bath computed from LE, GLE and MD simulations.}
	\label{fig:chain}
\end{figure}

Having established a very good performance for a single-particle dynamics, as
the next step in testing our proposed MLP-enriched GLE formalism we look at
the much more complex dynamics of a colloidal chain consisting of $N=20$
particles immersed in a thermal bath.
Particle chains are quite often used as prototypical systems to model
polymers.
In this context a widely used observable quantity for polymer characterization
is the gyration radius, $R_G=\sqrt{\frac{1}{N} \sum_{k=1}^N
	(\mathbf{r}_k-\mathbf{r}_{\text{CM} })^2}$, where $\mathbf{r}_{k}$ and
$\mathbf{r}_{\text{CM} }$ are the position vectors of the $k$-th particle and
the center of mass of the chain, respectively.
The increase in complexity of the ``target particle" (a coarse-grained
object, an aggregate of particles) brings about many complications when
trying to derive an appropriate coarse-grained dynamical equation.
We again rely upon the existence of a GLE which describes the time evolution
of the observable, in this case $R_G$, and which needs to be trained with
observed data, from MD in this particular case.
For the MD simulations, we again make use of an LJ potential $v_{LJ}$ to
model pairwise non-bonded interactions amongst chain and bath particles.
The chain particle interactions are given by the multi-body Dreiding
potential\cite{Mayo1990} adopted in several studies (e.g.
Ref.~\cite{Hossain2010}) to study polymer-chains deformations including
proteins in solution:
\begin{align}
v(\mathbf{r}_{i,j,k,l})=v_{\text{LJ}}(\mathbf{r}_{ij})+ v_{H}(\mathbf{r}_{ij})+ v_{\theta}(\mathbf{r}_{ijk})+v_{\phi}(\mathbf{r}_{ijkl}),
\label{eq:chain_potential}
\end{align}
where $v_{H}(\mathbf{r}_{ij})=k_H (\mathbf{r}_{ij} - \mathbf{r}_0)^2$,
$v_{\theta}(r_{ijk})= k_\theta (\theta_{ijk} - \theta_0)^2$ and
$v_{\phi}(r_{ijkl})= k_\phi (1+ \cos(2\phi_{ijkl}))$ account for linear,
angular and dihedral bonds, respectively (see Supplementary Information).
The bath has the same characteristics ($\rho_b=\frac{699}{1000}$, and
$T=1.0$) as in the LDL introduced in the previous section.
This choice, together with the assumption that the potential of mean force
acting amongst the chain particles is approximately equal to
$v(\mathbf{r}_{i,j,k,l})$, allows us to use the same memory kernel obtained
for the single particle (see Fig.~\ref{fig:bath_theta}).
With the MLP-approximated memory kernel fed into the GLE, we proceed with the
simulation of the $R_G$ dynamics, as we did for the single-particle case.
The results of such simulations are used to compute and compare the
gyration-radius autocorrelation\cite{Bishop1979,Dimitrov2008}, $C_{R_G}=
\left[ \langle R_G^2(t) R_G^2(0) \rangle- \langle R_G^2(0) \rangle^2 \right]
/ \left[ \langle R_G^4(0)  \rangle - \langle R_G^2(0) \rangle^2 \right]
\label{eq:Gyration_correlation}$, extracted from the different (G)LE
approximations (i.e., LE, GLE1 and GLE2) against the exact results obtained
from MD.
As can be readily checked in Fig. \ref{fig:chain}, in this case the GLE
enriched with a one-neuron MLP already is able to accurately reproduce the
bath effects on the chain, and clearly outperforms the commonly used
Markovian approximation.
The GLE2 model on the other hand seems to bring some regularization to the
autocorrelation, making it smoother than the one obtained from GLE1, although
both models seem to yield similar results.
Evidently, even with the simplest MLP model we are able to capture the
essence of the MD simulation, enabling us to carry out simulations with
MD-like quality at almost no computational cost.
The ease, but also convenience of our formalism, is indeed one of the most
remarkable facts that should be highlighted.

\subsection{Modelling global temperature}

\begin{figure}[t]
	\centering
	\begin{minipage}{0.4\textwidth}
		\flushleft (a)\\
		\centering
		\includegraphics[width=\textwidth]{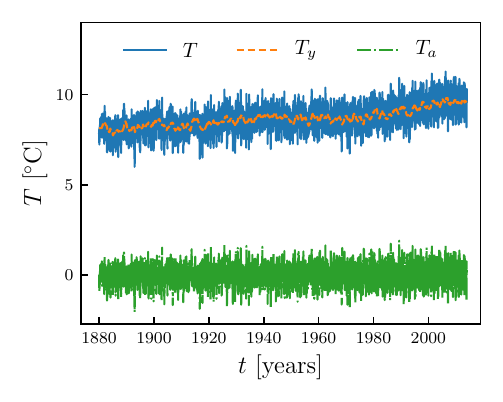}
	\end{minipage}
	\begin{minipage}{0.4\textwidth}
		\flushleft (b)\\
		\centering
		\includegraphics[width=\textwidth]{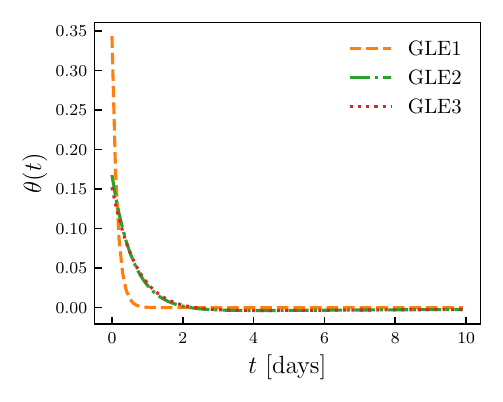}
	\end{minipage}
	\begin{minipage}{0.4\textwidth}
		\flushleft (c)\\
		\centering
		\includegraphics[width=\textwidth]{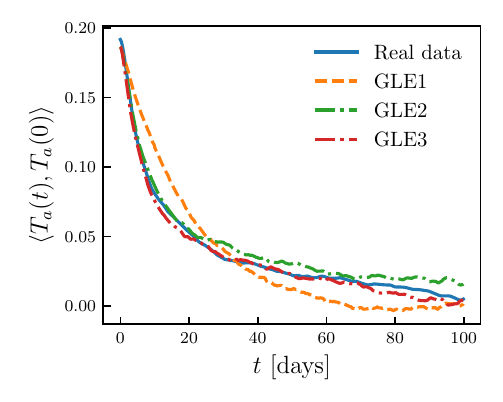}
	\end{minipage}
	\caption{(a) Global temperature  $T(t)$,  annual moving average temperature $T_y(t)$ and the daily anomaly $T_a(t)=T(t)-T_y(t)$ between 1880 and 2014. (b) Memory kernel approximations computed through MLP with 1,2 and 3 neurons in the hidden layer and (c) corresponding time correlations obtained from real data and GLE simulations.}
	\label{fig:temp}
\end{figure}

To illustrate now the versatility of our formalism we take a step away from
our familiar particle dynamics examples in the search for phenomena and
systems of general interest to look at.
Our first stop is the forecasting of the earth-global temperature. Public
attention aside, this is an interesting problem to tackle given that over the
last few decades several stochastic Markovian models have been proposed to
forecast global temperature dynamics, e.g.
Refs~\cite{Alaton2002,Moreles2016}.
GLEs can be viewed as natural generalization of such models, although the
memory kernel cannot be rigorously from first principles as in the case of
prototypical statistical-physical problems. And the natural question of
course here is what do GLEs have to do with climate. However, let us not
forget the basic structure of GLEs: the dynamic variation of an observable
quantity, a force, a drift, a term reflecting the history of the system and
noise. The basic ingredients of most, if not all stochastic systems. But in
the majority of the cases first-principles dynamical laws for stochastic
systems -- the ``\emph{reductionist approach}" as was referred to by the
authors in Ref.~\cite{Koren2011} -- are just not possible. In such cases the
``\emph{complementary systems-based approach}" -- again, referred to by the
authors in Ref.~\cite{Koren2011} -- is the only way out. But stipulating a
model is not the end of the road, rather the beginning. The model needs to be
properly parametrized and GLEs are not an exception. It is only then that
they are capable of describing non-Markovian Gaussian processes and, thus,
model general stochastic time series.
%
Here we demonstrate that an MLP-enriched GLE is able to accurately describe
the daily global-temperature fluctuations with respect to a properly chosen
moving average (although our methodology can be also employed to model local
temperature dynamics).
Consider the daily land-average global temperature $T(t)$ measured during the
period 1880-2014, published by Berkeley
Earth\cite{database-Berkeley,Rohde2013}.
Despite the local temperature showing cyclical trends in short periods (e.g.
due to to season changes), $T(t)$ does not exhibit a significant seasonal
behavior, this being the result of the energy balance between solar and earth
radiations\cite{Friis-Christensen1991}.
Nevertheless, $T(t)$ reveals non-stationarity features due to a long-period
increasing trend related to global warming, as observed in
Fig.~\ref{fig:temp}(a).
We first compute the long term dynamics $T_y(t)$ as a yearly moving average.
We then define the observable of interest as $T_a(t)=T(t)-T_y(t)$, so that
the corresponding time series is stationary (see Supplementary Information).
Hence, our proposal is to model the time evolution of $T_a(t)$ with the GLE $\partial_t T_a(t) = -\int_{0}^{t} \theta (t-\tau) T_a (\tau) d \tau +
R(t)$.
In fact, this is a generalization of the Markovian model for weather
derivatives proposed in Ref.~\cite{Alaton2002}.
The different approximations of the memory kernel extracted from our
MLP-based method are shown In Fig.~\ref{fig:temp}(b), while
Fig.~\ref{fig:temp}(c) shows the corresponding correlation functions.
We observe an excellent agreement between the correlations obtained with the
GLE dynamics and the real-world data, especially when three neurons are
adopted in the hidden layer (GLE3).
Matching then the relaxation times of the memory kernel ($\sim$ days) with
the characteristic time of the variable $T_y$ ($\sim$ years), we can obtain
the evolution of $T(t)$ as a sum of a Markovian yearly (long term)
contribution and a non-Markovian daily (short term) contribution, namely:
\begin{align}
\partial_t T(t) =  \left( \partial_t  + \theta_c \right) T_y(t)  -   \int_{0}^{t} \theta (t-\tau) T (\tau) d \tau + R(t) ,
\label{eq:temp}
\end{align}
where the constant is given by $\theta_c=\int_{0}^{t} \theta (\tau)$.
Equation~\eqref{eq:temp}, originating directly from data, reflects the main
features of global-temperature multi-scale dynamics and, thus, gives clear
insights into current questions regarding, for instance, global warming.
In fact, Eq.~\eqref{eq:temp} can be used to distinguish between long-term temperature trends and short term fluctuations.

\subsection{A stock market model: The Nikkei index}

\begin{figure*}[h!]
	\centering
	\begin{minipage}{0.4\textwidth}
		\flushleft (a)\\
		\centering
		\includegraphics[width=1\textwidth]{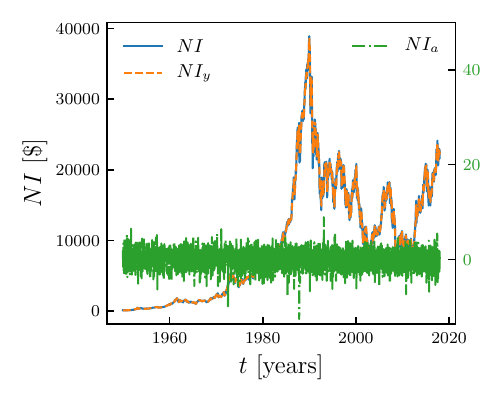}
	\end{minipage}
	\begin{minipage}{0.4\textwidth}
		\flushleft (b)\\
		\centering
		\includegraphics[width=1\textwidth]{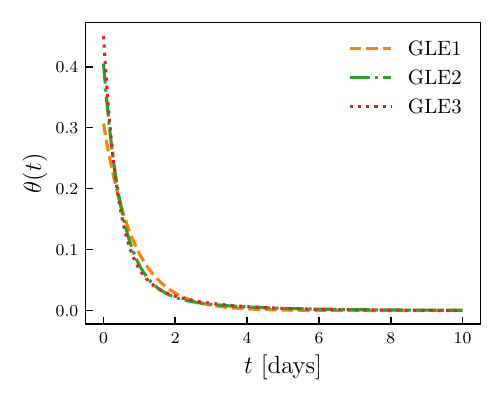}
	\end{minipage}
	\begin{minipage}{0.4\textwidth}
		\flushleft (c)\\
		\centering
		\includegraphics[width=1\textwidth]{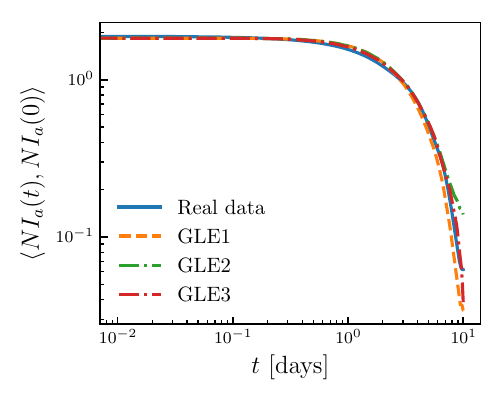}
	\end{minipage}
	\begin{minipage}{0.4\textwidth}
		\flushleft (d)\\
		\centering
		\includegraphics[width=1\textwidth]{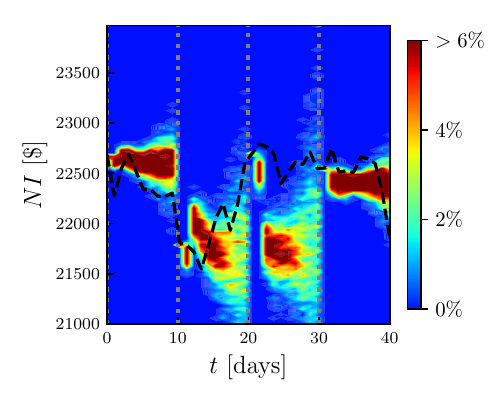}
	\end{minipage}
	\caption{
		(a) Daily close price of Nikkei index $\text{NI}(t)$,  moving average index $\text{NI}_y(t)$ computed over a window of ten days preceding the time $t$ and the normalized index $\text{NI}_a (t)$ between 1949 and 2018.
		(b) Memory kernel approximations computed through MLP with 1,2 and 3 neurons in the hidden layer and (c) the corresponding time correlations obtained from real data and GLE simulation.
		(d) Comparison between predicted probability distribution (color-map) and actual market data (dashed black line).
		Dotted lines in gray delineate the 10 days long investment windows.
	}
	\label{fig:Nikkei}
\end{figure*}

As expected, stochastic models have widely been employed to gain insight into
financial instruments, such as bonds and stock
prices\cite{Takahashi1996,Lee2002}. This is driven by the need that financial
operations, such as financial risk management and portfolio optimization,
require accurate predictions of markets dynamics to maximize profits.
However, the majority of the models used in finance rely on Markovian
assumptions, which can potentially introduce inaccuracies.
Moreover, alternative standard approaches to model directly data series have a range of applicability limited by the training dataset. Here we use physics informed priors to integrate our experimental intuition of the phenomenon.
Here we show that
our methodology can overcome such limitations.
As a case study here we adopt the GLE to
model the daily price of the Japanese financial index Nikkei, $\text{NI}(t)$,
between May 1949 and May 2018\cite{database-Nikkei}. As with many other
financial instruments, $\text{NI}(t)$ exhibits non-stationary behavior in
both mean and variance. Assuming a local equilibrium approximation in the
Nikkei trend (as we would do in Thermodynamics for a phenomenon at hand), we
build an observable defined as $\text{NI}_a (t) = \left[ \text{NI}(t) -
\text{NI}_y (t) \right] / \sigma_y (t) $, with $\text{NI}_y (t)$ and
$\sigma_y (t)$ being respectively a moving average and a moving standard
deviation computed over a period $[t-y,t-1]$, respectively. The parameter $y$
is then selected so as to obtain a stationary $\text{NI}_a (t)$; we find
$y=10$ days to be the optimal value (details are given in the Supplementary Information. Hence, we model the normalized stock price $\text{NI}_a(t)$ with
the following non-Markovian model: $\partial_t \text{NI}_a (t) =
-\int_{0}^{t} \theta (t-\tau) \ \text{NI}_a(\tau) d \tau + R(t)
\label{eq:nikkei}$. In Fig.~\ref{fig:Nikkei}(a), we report the observable
$\text{NI}_a (t)$ which exhibits a stationary Gaussian behavior (as verified
in the Supplementary Information) and, thus, confirms our assumption of local
equilibrium dynamics.
Figures~\ref{fig:Nikkei}(b)-(c) show various degrees of approximations
obtained with our framework and the corresponding correlation functions. In
contrast with the global temperature trend, $\text{NI}_a(t)$ do not exhibit a
clear time-scale separation between the memory kernel and autocorrelation
decay. The GLE dynamics shows a growing accuracy in representing the real
data when the number of neurons in the hidden layer increases. As a matter of
fact, already with the third-order approximation we are able to reproduce the
correlation decay with a maximum relative error of order $10^{-2}$. The
proposed GLE equation, parameterized with an MLP equipped with 3 neurons, is
then employed in a comparison between the predicted probability distribution
and actual market data for four time windows, each ten market days long,
between Jun 2018 and Aug 2018 (Fig.~\ref{fig:Nikkei}(d)). It is clear that
our model is able not only to predict most of the actual market trend, but,
crucially, to provide quite accurate information on the local variance of the
trend, thus opening the way to optimizing risk management in short-term
($\sim$ weekly) investments.

\section{Conclusions}
We have introduced a novel methodology to decipher the
analytically-intractable GLE dynamics.
The basis of our framework is (a) stipulate GLEs as the fundamental
underlying model for real-world stochastic systems, and (b) enriching GLEs
with elements of machine learning such as neural networks. 
The universal approximation theorem guarantees the
general applicability of our methodology.
We have demonstrated that our machine-learning-enriched GLE is both
accurate and efficient. 
But also robust when it comes to dealing with data
affected by natural fluctuations which is typically the case with real-system
data sets.

	For convenience of the reader, the main steps required to apply the proposed methodology are listed below:
	\begin{itemize}
		\item Compute the matrices $\mathbf{g}(t)$ and $\mathbf{h}(t)$ (Sect.\ref{method}) from a historical data series sampling the dynamical evolution of the observable of interest.
		\item Estimate the memory kernel function by means of a MLP with the structure described in Sect.~\ref{MLP}
		\item Model the dynamical evolution of the system by employing a GLE embedded with the memory kernel computed in the previous step. For the integration in time of the GLE, one could follow the extended variables framework discussed in Sect.\ref{GLE_integration}
	\end{itemize}

We have successfully tested our methodology against several prototypical
examples: from standard problems like a single colloidal particle and
particle chains in a bath, to climatology and finance.
In all cases, we found
excellent agreement between the actual and the approximated dynamics of the
observables under consideration.
Thus, coupling machine learning with a general equation of statistical
mechanics, namely GLE, offers an attractive and versatile computational
toolbox opening the door to a new way of modelling and understanding
stochastic systems and, more general, doing statistical mechanics.
Future developments include relaxing the Markovian approximation in
dynamic-density functional theory and fluctuating
hydrodynamics\cite{Miguel2017,Ben2012} but also adopting MLPs equipped with
complex-valued exponential functions aiming to approximate oscillatory memory
kernels.

\section*{Supplementary Information}

\appendix

\begin{figure*}[h!]
	\centering
	\begin{minipage}{0.4\textwidth}
		\flushleft (a)\\
		\centering	
		\includegraphics[width=\textwidth]{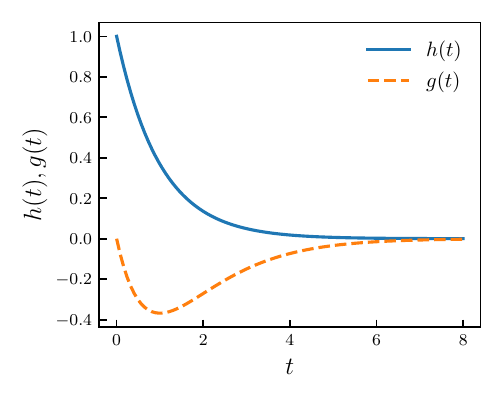}
	\end{minipage}
	\begin{minipage}{0.4\textwidth}
		\flushleft (b)\\
		\centering	
		\includegraphics[width=\textwidth]{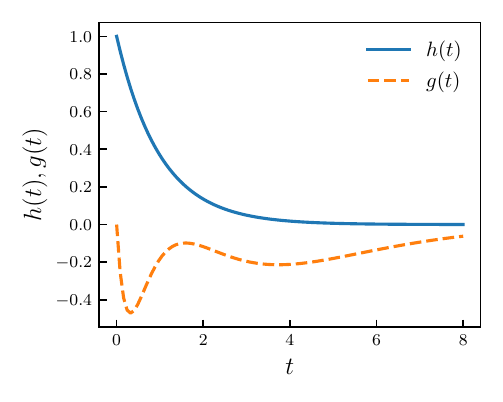}
	\end{minipage}
	\begin{minipage}{0.4\textwidth}
		\flushleft (c)\\
		\centering	
		\includegraphics[width=\textwidth]{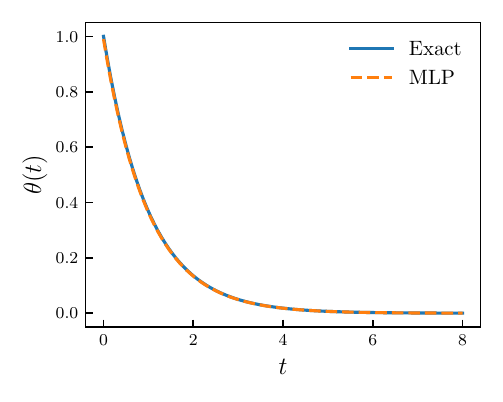}
	\end{minipage}
	\begin{minipage}{0.4\textwidth}
		\flushleft (d)\\
		\centering	
		\includegraphics[width=\textwidth]{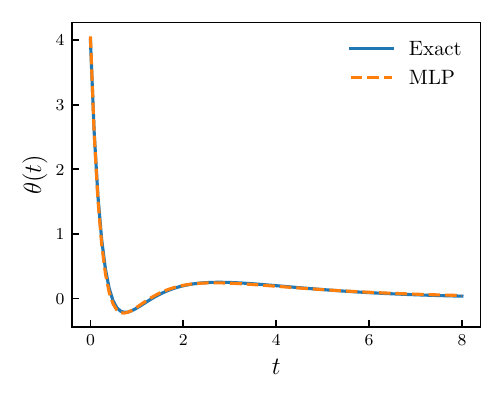}
	\end{minipage}
	\begin{minipage}{0.4\textwidth}
		\flushleft (e)\\
		\centering	
		\includegraphics[width=\textwidth]{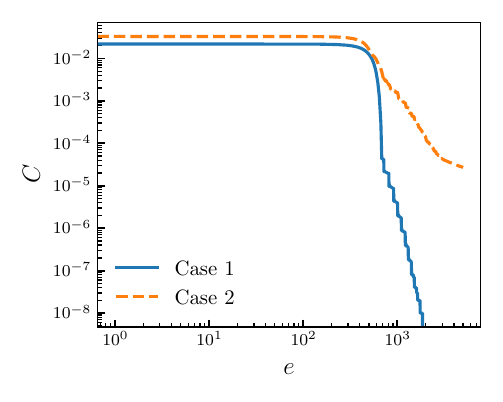}
	\end{minipage}
	\begin{minipage}{0.4\textwidth}
		\flushleft (f)\\
		\centering	
		\includegraphics[width=\textwidth]{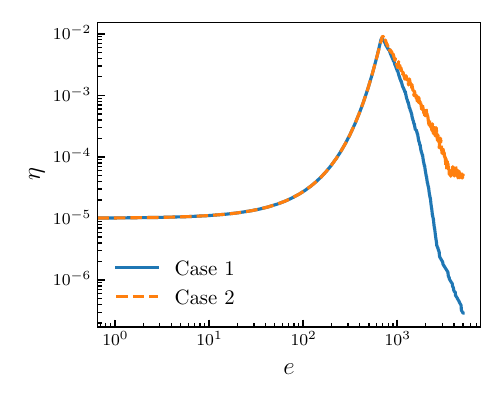}
	\end{minipage}
	\caption{
		Two representative cases (discussed in the main text) to test our methodology: Case 1 in (a-c) and case 2 in (b-d).
		The functions $h(t)$ and $g(t)$ (a-b), discretized at $800$ points, are provided to the MLP.
		The comparison between the memory kernel $\theta$ computed numerically with our MLP and the exact one is given in (c-d).
		In (e-f) we show the cost function and learning rate for the two analyzed scenarios.
		In both cases, the numerical approximation is obtained with an MLP trained for 5,000 epochs.
	}
	\label{fig:memory_kernel_exact}
\end{figure*}

\section{Multi-layer perceptron structure: additional details and results}

\begin{figure}[ht]
	\centering	
	\begin{minipage}{0.4\textwidth}
		\centering	
		\includegraphics[width=\textwidth]{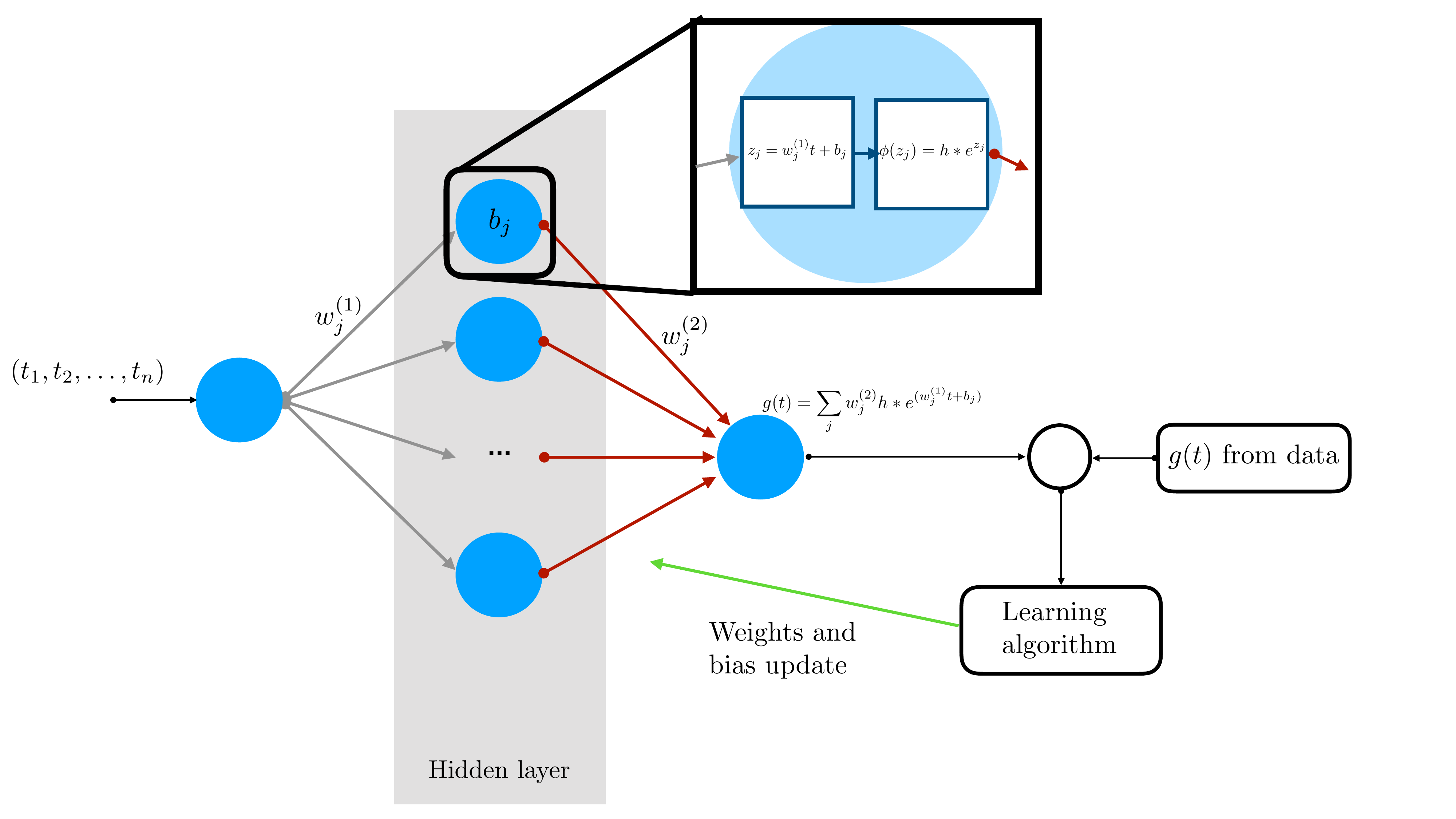}
	\end{minipage}
	\caption{Representation of a MLP structure and the learning process}
	\label{fig:highD_nn}
\end{figure}

The structure of our MLPs is depicted in Fig.~\ref{fig:highD_nn}.
The network training process is based on the updates of weights and biases.
For the quadratic cost function the updates are defined as:
\begin{itemize}
	\item error at the output layer: $\delta^{(3)}=\frac{\partial C}{\partial z^{(3)}}= \frac{\partial C}{\partial a^{(3)}} \frac{\partial a^{(3)}}{\partial z^{(3)}}= \frac{1}{N_t} \sum_{t_j}^{N_t} \left( y_{j}(t_j)- a_j^{L}(t_j) \right)$;
	\item error of the neuron j at the layer l: $\delta_j^l=\frac{\partial C}{\partial z_j^l}= \sum_{k}^{}  w_{j,i}^{l+1} \delta_{k}^{l+1} \frac{\partial \phi(z_j^l)}{ \partial z_j^l}$;
	\item update of weight: $\Delta w_{j,i}^l =  -\eta a_k^{l-1} \frac{\partial C}{\partial z_j^l}$;
	\item update of bias: $\Delta b_j^l = -\eta \frac{\partial C}{\partial z_j^l}$.
\end{itemize}

As a preliminary test of our approach, we consider three simple functions
$h(t)$, $\theta(t)$ and $g(t)$ for which it can be shown analytically that
$g(t)= -\int_{0}^{t} \theta (t-\tau) h (\tau) d \tau$.
Given $h(t)$ and $g(t)$, an approximation of $\theta(t)$ is then computed
with our methodology and is compared with the analytical $\theta(t)$.
Two tests with different sets of functions are reported here.
The functions used for the first test are:
\begin{align}
h(t) = e^{-t} , \quad \theta(t) = e^{-t} , \quad g(t)  = - t e^{-t} .
\label{eq:perceptron_test1}
\end{align}
In this test, because of the single exponential form
of $\theta$, an MLP with a single neuron in the hidden layer is used, namely
$N_n=1$.

The functions adopted for the second test are:
\begin{align}
\begin{split}
&h(t) = e^{-t} , \quad \theta(t) = 6 e^{-4t} -4 e^{-t}+2 e^{-t/2} , \\
&g(t)  = - \left( 2 e^{-t} -2 e^{-4t} -4 t e^{-t}+ 4 e^{3 t/2}-4 e^{-t} \right) .
\label{eq:perceptron_test2}
\end{split}
\end{align}
For this latter example, we impose $N_n=3$ neurons in the hidden layer.

Figures~\ref{fig:memory_kernel_exact}(a,b) report $h(t)$ and $g(t)$ provided
as input to the MLP for both tests.
The comparisons between numerical approximations and analytical $\theta$ reported in Fig.~\ref{fig:memory_kernel_exact}(c-d) clearly shows the accuracy of our methodology.
The behaviors of cost function and learning rate during the learning process
for both tests are also shown in Fig.~\ref{fig:memory_kernel_exact}(e-f).

\begin{figure}[ht]
	\centering
	\begin{minipage}{0.4\textwidth}
		\flushleft (a)\\
		\centering	
		\includegraphics[width=\textwidth]{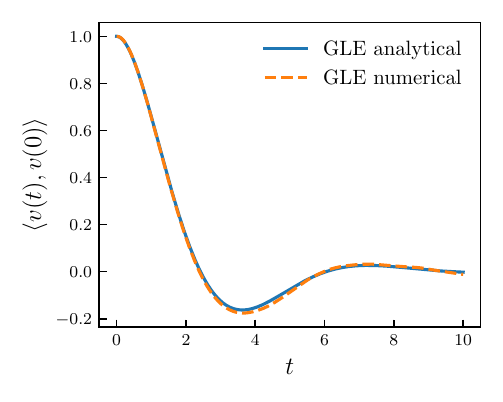}
	\end{minipage}
	\begin{minipage}{0.4\textwidth}
		\flushleft (b)\\
		\centering	
		\includegraphics[width=\textwidth]{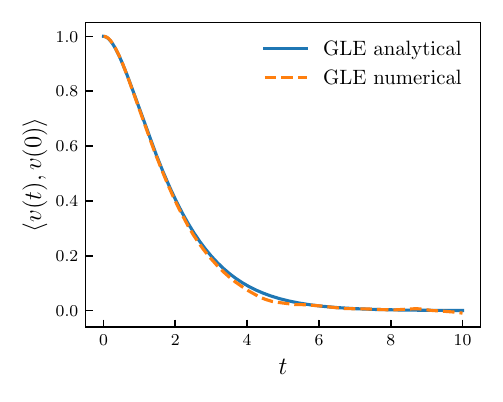}
	\end{minipage}
	\begin{minipage}{0.4\textwidth}
		\flushleft (c)\\
		\centering	
		\includegraphics[width=\textwidth]{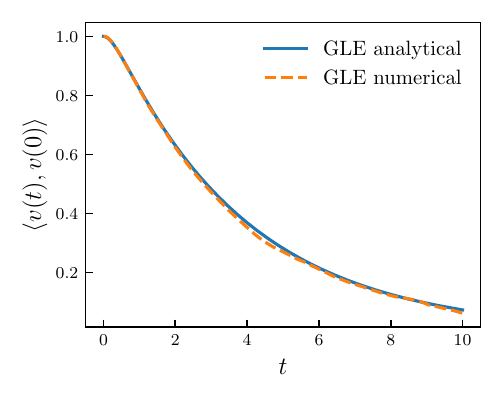}
	\end{minipage}
	\caption{Comparison between numerical and analytical time correlation computed over $10^4$ independent trajectories for a GLE with memory kernel in the form of a single
		exponential function~\cite{Baczewski2013}.
		The correlation is computed in the under-damped limit with A=1 and B=1 (a), in the damped case with A=1 and B=-2 (b),  and in the over-damped limit with A=1 and B=-4 (c).
		In all cases the temperature is set to $T=1$.
	}
	\label{fig:autv}
\end{figure}

\begin{figure*}[h!]
	\centering
	\begin{minipage}{0.4\textwidth}
		\flushleft (a)\\
		\centering	
		\includegraphics[width=\textwidth]{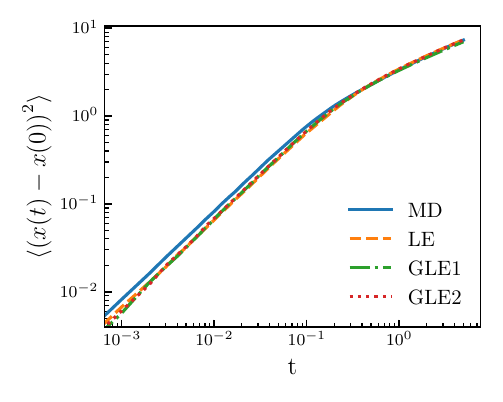}
	\end{minipage}
	\begin{minipage}{0.4\textwidth}
		\flushleft (b)\\
		\centering	
		\includegraphics[width=\textwidth]{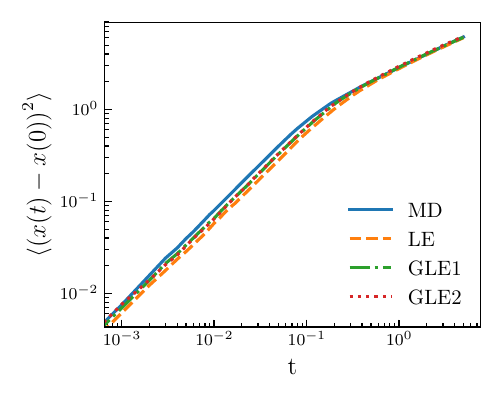}
	\end{minipage}
	\begin{minipage}{0.4\textwidth}
		\flushleft (c)\\
		\centering	
		\includegraphics[width=\textwidth]{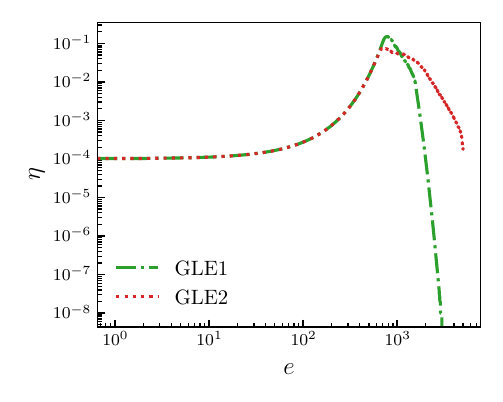}
	\end{minipage}
	\begin{minipage}{0.4\textwidth}
		\flushleft (d)\\
		\centering	
		\includegraphics[width=\textwidth]{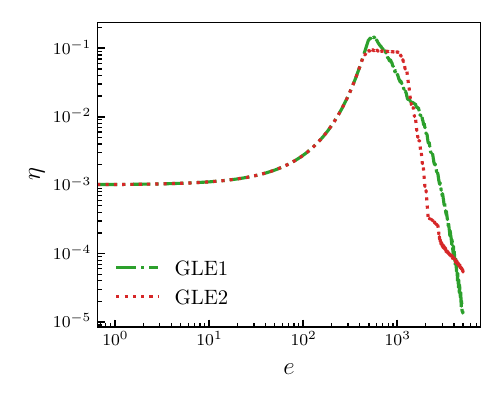}
	\end{minipage}
	\begin{minipage}{0.4\textwidth}
		\flushleft (e)\\
		\centering	
		\includegraphics[width=\textwidth]{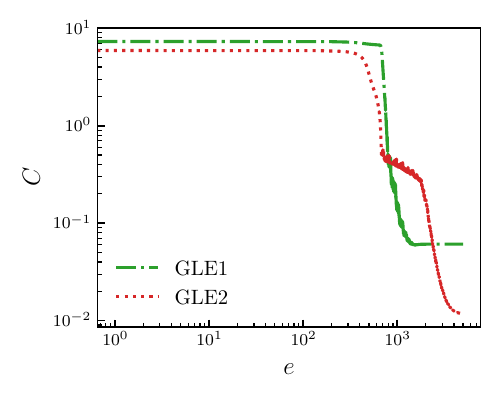}
	\end{minipage}
	\begin{minipage}{0.4\textwidth}
		\flushleft (f)\\
		\centering	
		\includegraphics[width=\textwidth]{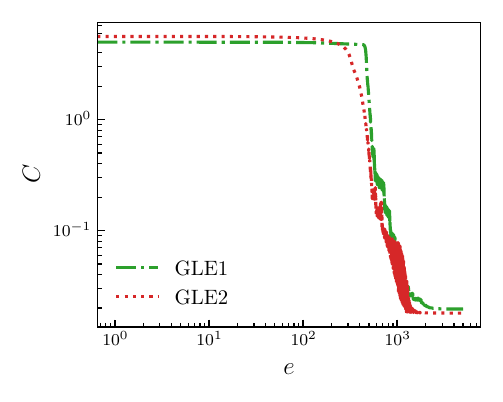}
	\end{minipage}
	\caption{Comparison of the mean-square displacement MSD computed with the MLP and with MD, for both LDL (a) and HDL (b) cases.
		GLE1 and GLE2 refer to the memory kernel approximations obtained with 1 and 2 neurons in the hidden layer, respectively.
	}
	\label{fig:bath_theta1}
\end{figure*}

\begin{figure*}[!t]
	\centering
	\begin{minipage}{0.3\textwidth}
		\flushleft (a)\\
		\centering	
		$t=1$
		\includegraphics[width=\textwidth]{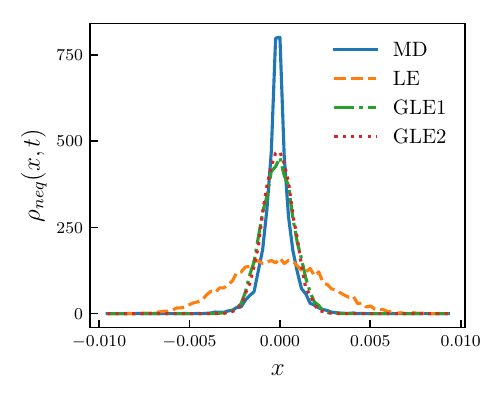}
	\end{minipage}
	\begin{minipage}{0.3\textwidth}
		\flushleft (b) \\
		\centering	
		$t=1$
		\includegraphics[width=\textwidth]{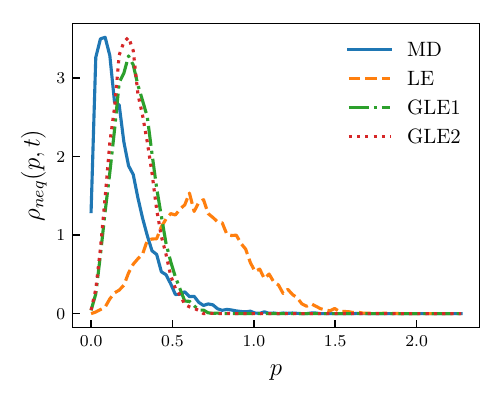}
	\end{minipage}
	
	\begin{minipage}{0.3\textwidth}
		\flushleft (c)\\	
		\centering	
		$t=5$
		\includegraphics[width=\textwidth]{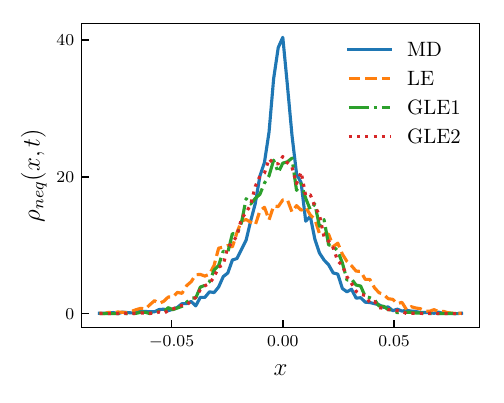}
	\end{minipage}
	\begin{minipage}{0.3\textwidth}
		\flushleft (d)\\	
		\centering	
		$t=5$
		\includegraphics[width=\textwidth]{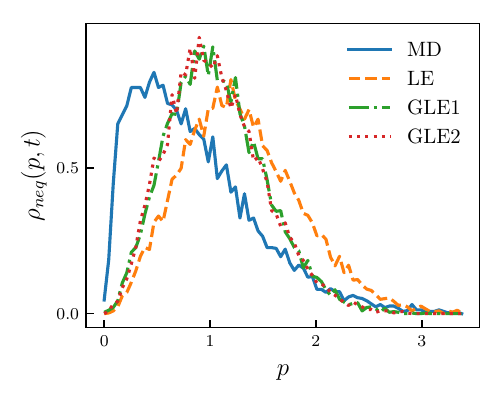}
	\end{minipage}
	
	\begin{minipage}{0.3\textwidth}
		\flushleft (e)\\	
		\centering	
		$t=30$
		\includegraphics[width=\textwidth]{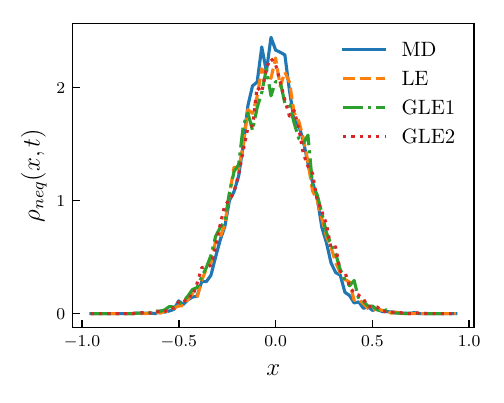}
	\end{minipage}
	\begin{minipage}{0.3\textwidth}
		\flushleft (f)\\	
		\centering	
		$t=30$
		\includegraphics[width=\textwidth]{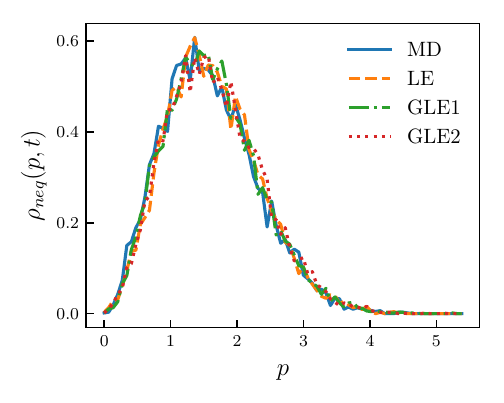}
	\end{minipage}
	
	\begin{minipage}{0.3\textwidth}
		\flushleft (g)\\	
		\centering	
		\includegraphics[width=\textwidth]{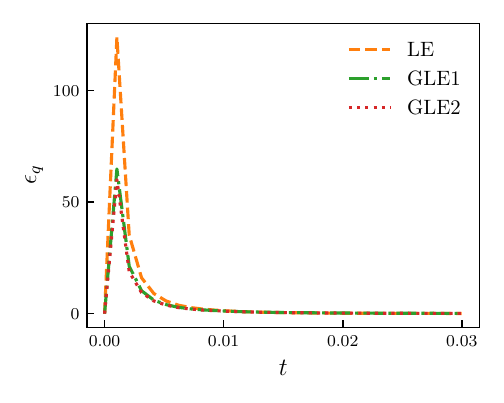}
	\end{minipage}
	\begin{minipage}{0.3\textwidth}
		\flushleft (h)\\	
		\centering	
		\includegraphics[width=\textwidth]{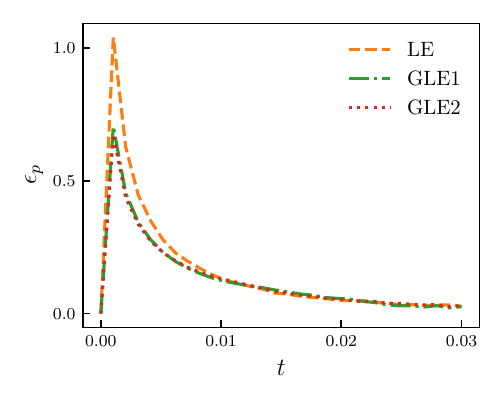}
	\end{minipage}
	\caption{
		Relaxation dynamics of position (a-c-e) and momentum (b-d-f) PDF from Dirac delta to equilibrium computed with MD, LE and GLE over $10^4$ trajectories.
		Corresponding mean square error of position (g) and momentum (h) PDF in time.}
	\label{fig:bath_nneq_pdf}
\end{figure*}

\section{Numerical methods: GLE integration}


\paragraph{Convolution decomposition}

As mentioned in the main text, the convolution term is written as a sum of the additional variable vectors $\mathbf{Z}_{k}(t)$, with $Z_{k,i}(t)= \int_{0}^{t} A_{k,ij} e^{B_{k,ij} (t-\tau)} \boldsymbol{ \mathcal{O} }_j (\tau) d \tau $.
Applying Leibniz's integral rule, and taking advantage of the symmetry of the
matrices $ \mathbf{B}_k$, it follows:
\begin{align}
\dot{\mathbf{Z}}_{k}(t)= \mathbf{B}_k \mathbf{Z}_{k}(t) - \mathbf{A}_k \boldsymbol{ \mathcal{O} }(t) .
\end{align}
Hence, the original GLE is rewritten in the equivalent form:
\begin{align}
\begin{cases}
\partial_t \boldsymbol{ \mathcal{O} }(t) = \mathcal{P} \mathcal{L} \boldsymbol{ \mathcal{O} } - \sum_k \mathbf{Z}_{k}(t) + \mathbf{R}(t) , \\
\partial_t \mathbf{Z}_{k}(t)= \mathbf{B}_k \mathbf{Z}_{k}(t) + \mathbf{A}_k \boldsymbol{ \mathcal{O} }(t) .
\end{cases}
\end{align}

\paragraph{Random force decomposition}

We now turn to the theoretical derivation of the random force decomposition
for a general real tensor function $\boldsymbol{\theta}(t)$.
It is worth noticing that such formulation is valid for any form of the memory kernel, not just exponential ones.
First, let us remark that, because of the symmetry between $t$ and $t'$ in
the fluctuation-dissipation theorem, $\boldsymbol{\theta}(t)$ is an even
function of time, i.e. $\boldsymbol{\theta}(t)=\boldsymbol{\theta}(-t)$.
Let us now define the Fourier transform of $\boldsymbol{\theta}(t)$ as
$\boldsymbol{\widetilde{\theta} } (\omega) = \int_{-\infty}^{-\infty}
\boldsymbol{\theta}(t) e^{ -i \omega \tau } d t$.
Since $\boldsymbol{\theta}(t)$ is real and even in time,
$\boldsymbol{\widetilde{\theta} } (\omega)$ is also real and even for real
$\omega$.
It follows that both zeros and singular points of
$\boldsymbol{\widetilde{\theta} } (\omega)$ are symmetric with respect to
both real and imaginary axes in the $\omega$-plane.
We then introduce the function $\boldsymbol{\widetilde{\chi}}(\omega)$ given
by
\begin{align}
\boldsymbol{\widetilde{\chi}}(\omega) =  \sum_{k} -i   \left(\omega \mathbf{I} + i \mathbf{B'}_k\right)^{-1}  \mathbf{b}_k ,
\label{supp:eq:g}
\end{align}
where the real matrices $\mathbf{b}_k$ and $\mathbf{B'}_k$ are such that:
\begin{align}
\boldsymbol{\widetilde{\theta}}(\omega) \langle \boldsymbol{ \mathcal{O} } , \boldsymbol{ \mathcal{O} } \rangle = 2 \boldsymbol{\widetilde{\chi}}(\omega) \boldsymbol{\widetilde{\chi}}^T(-\omega) ,
\label{supp:eq:theta-g}
\end{align}
and the singular points of $\boldsymbol{\widetilde{\chi}}^{-1}(\omega)$ lie in the lower-half complex $\omega$-plane.
Moreover, we define the two matrices:
\begin{align}
\boldsymbol{\widetilde{\zeta}}(\omega) =  \boldsymbol{\widetilde{\chi}}^{-1}(\omega)= \sum_{k}   \left(\omega \mathbf{I} + i \mathbf{B'}_k\right) \left(-i \mathbf{b}_k\right)^{-1} ,
\label{supp:eq:h}
\end{align}
and
\begin{align}
\mathbf{\widetilde{k}}_k(\omega) =  -i  \left(\omega \mathbf{I} + i \mathbf{B'}_k\right)^{-1}  \mathbf{b}_k \boldsymbol{\widetilde{\zeta}}(\omega) ,
\label{supp:eq:k}
\end{align}
and we denote their Fourier inverse transform with
$\boldsymbol{\zeta}(t)$ and $\mathbf{k}_k(t)$.
Combining Eqs~\eqref{supp:eq:g}, \eqref{supp:eq:h} and \eqref{supp:eq:k}, it
follows that:
\begin{align}
\sum_k  \mathbf{\widetilde{k}}_k(\omega) =  \mathbf{I}
\label{supp:eq:k1}
\end{align}
or, equivalently,
\begin{align}
\sum_k \mathbf{k}_k (t) = \mathbf{I} \delta(t).
\label{supp:eq:k2}
\end{align}
Moreover Eq.~\eqref{supp:eq:k} can be rewritten as $\left( i \omega \mathbf{I} - \mathbf{B'}_k\right) \mathbf{\widetilde{k}}_k(\omega) =   \mathbf{b}_k \boldsymbol{\widetilde{\zeta}}(\omega)$, that in the time domain gives:
\begin{align}
\frac{d}{d t}\mathbf{k}_k (t) - \mathbf{B'}_k  \mathbf{k}_k (t) = \mathbf{b}_k . \boldsymbol{\zeta}(t)
\label{supp:eq:k3}
\end{align}
Finally, the following vector variables are introduced:
\begin{align}
\boldsymbol{\xi}(t) = \int_{0}^{+\infty} \boldsymbol{\zeta} (t-t')  \mathbf{R} (t') d t' ,
\label{supp:eq:eta}
\end{align}
and
\begin{align}
\mathbf{R}_k(t) = \int_{0}^{+\infty} \mathbf{k}_k (t-t')  \mathbf{R} (t') d t' .
\label{supp:eq:xi}
\end{align}
From Eq.~\eqref{supp:eq:xi} and Eq.~\eqref{supp:eq:k2} it follows that:
\begin{align}
\sum_k \mathbf{R}_k (t) =  \mathbf{R} (t) ,
\label{supp:eq:xi1}
\end{align}
while, combining Eq.~\eqref{supp:eq:xi} and Eq.~\eqref{supp:eq:k3}
\begin{align}
\frac{d}{d t} \mathbf{R}_k(t)=  \mathbf{B'}_k  \mathbf{R}_k(t) + \mathbf{b}_k \boldsymbol{\xi}(t).
\label{supp:eq:xi2}
\end{align}
Equations~$\left(\ref{supp:eq:xi1}-\ref{supp:eq:xi2}\right)$ are our main result here.
They allow to express the correlated noise of the original GLE as a function
of white noise $\boldsymbol{\xi}(t)$.

In what follows, we discuss the properties of the stochastic process
$\boldsymbol{\xi}(t)$. First, since all the singularities of
$\boldsymbol{\widetilde{\zeta}}(\omega)
=\boldsymbol{\widetilde{\chi}}^{-1}(\omega)$ lie in the lower-half complex
$\omega$-plane, it follows that for $\tau>0$ we have:
\begin{align}
\begin{split}
\boldsymbol{\zeta} (\tau) & = \frac{1}{2 \pi} \int_{-\infty}^{\infty} d \omega   \boldsymbol{\widetilde{\zeta}}(\omega) e^{i \omega \tau} \\
& = \lim\limits_{a\rightarrow \infty} \frac{1}{2 \pi} \int_{-a}^{a} d \omega  \boldsymbol{\widetilde{\zeta}}(\omega) e^{i \omega \tau} =\\
&=\frac{1}{2 \pi} \oint_{C^+} d \omega  \boldsymbol{\widetilde{\zeta}} (\omega) e^{i \omega \tau} \\
& - \lim\limits_{a\rightarrow \infty} \frac{1}{2 \pi} \int_{ \text{arc} (a \rightarrow -a)} d \omega  \boldsymbol{\widetilde{\zeta}}(\omega) e^{i \omega \tau} = \mathbf{0} ,
\label{supp:zeta_mean}
\end{split}
\end{align}
where $\oint_{C^+} d \omega$ indicates the integral over a closed contour
$C^+$ along the real line from $-a$ to $a$ and then along a circular arc at
$0$ from $a$ to $-a$ and in the upper half-plane, $\int_{ \text{arc} (a
	\rightarrow -a)} d \omega $.
Hence, for $t>0$ we can write
\begin{align}
\boldsymbol{\xi}(t) &= \int_{0}^{+\infty} \boldsymbol{\zeta} (t-t')  \mathbf{R} (t') d t' \nonumber \\
&=  \int_{-\infty}^{+\infty} \boldsymbol{\zeta} (t-t')  \mathbf{R} (t') d t' .
\label{supp:eq:xi3}
\end{align}
Thus, the correlation function of $\boldsymbol{\xi}(t)$ at $t_1$ and $t_2$ is given by:
\begin{align}
\begin{split}
\left\langle \boldsymbol{\xi}(t_1) \boldsymbol{\xi}^T(t_2) \right \rangle & =
\int_{-\infty}^{+\infty} dt_1' \int_{-\infty}^{+\infty} dt_2' \boldsymbol{\zeta} (t_1-t_1') \times \\
& \left\langle \mathbf{R} (t_1') \mathbf{R}^T (t_2') \right \rangle \boldsymbol{\zeta}^T (t_2-t_2')  = \\
& = \int_{-\infty}^{+\infty} dt_1' \int_{-\infty}^{+\infty} dt_2' \boldsymbol{\zeta} (t_1-t_1') \boldsymbol{\theta}(t_1'-t_2') \times \\
&\langle \boldsymbol{ \mathcal{O} } , \boldsymbol{ \mathcal{O} } \rangle \boldsymbol{\zeta}^T (t_2-t_2')  , \\
\label{supp:xi_corr}
\end{split}
\end{align}
where we used the fluctuation-dissipation theorem.
From the definition of Fourier transform of $\boldsymbol{\theta}$, it follows
\begin{align}
\begin{split}
\left\langle \boldsymbol{\xi}(t_1) \boldsymbol{\xi}^T(t_2) \right \rangle &= \int_{-\infty}^{+\infty} dt_1' \int_{-\infty}^{+\infty} dt_2' \boldsymbol{\zeta} (t_1-t_1') \times \\
& \frac{1}{2 \pi} \int_{-\infty}^{+\infty} d \omega \boldsymbol{\widetilde{\theta}}(\omega) e^{ i \omega (t_1'-t_2') }   \langle \boldsymbol{ \mathcal{O} } , \boldsymbol{ \mathcal{O} } \rangle \times \\ & \boldsymbol{\zeta}^T (t_2-t_2') = \\
&= \int_{-\infty}^{+\infty} dt_1' \int_{-\infty}^{+\infty} dt_2' \boldsymbol{\zeta} (t_1-t_1') \times \\
& \frac{1}{2 \pi} \int_{-\infty}^{+\infty} d \omega \boldsymbol{\widetilde{\theta}}(\omega) e^{ i \omega (t_1'-t_2') }    \langle \boldsymbol{ \mathcal{O} } , \boldsymbol{ \mathcal{O} } \rangle  \boldsymbol{\zeta}^T (t_2-t_2') \times \\
&e^{ -i \omega (t_1-t_2) } e^{ i \omega (t_1-t_2) }  = \\
&= \frac{1}{2 \pi}  \int_{-\infty}^{+\infty} d \omega   \int_{-\infty}^{+\infty} dt_1' \boldsymbol{\zeta} (t_1-t_1') \times \\
& e^{-i \omega (t_1-t_1') }  \boldsymbol{\widetilde{\theta}}(\omega) \langle \boldsymbol{ \mathcal{O} } , \boldsymbol{ \mathcal{O} } \rangle  \int_{-\infty}^{+\infty} dt_2'  \boldsymbol{\zeta}^T (t_2-t_2') \times \\
&e^{ i \omega (t_2'-t_2') } e^{ i \omega (t_1-t_2) } .
\label{supp:xi_corr1}
\end{split}
\end{align}
Applying the definition of Fourier transform of $\boldsymbol{\zeta}(t)$, and taking advantage of Eq.~\eqref{supp:eq:theta-g} and Eq.~\eqref{supp:eq:h}, we finally obtain:
\begin{align}
\begin{split}	
\left\langle \boldsymbol{\xi}(t_1) \boldsymbol{\xi}(t_2)^T \right \rangle  = & \frac{1}{2 \pi} \int_{-\infty}^{+\infty} d \omega  \boldsymbol{\widetilde{\zeta}}(\omega) \boldsymbol{\widetilde{\theta}}(\omega) \langle \boldsymbol{ \mathcal{O} } , \boldsymbol{ \mathcal{O} } \rangle \\
&  \boldsymbol{\widetilde{\zeta}}^T(-\omega) e^{ i \omega (t_1-t_2) }  = \\
= &  \frac{1}{2 \pi}  \int_{-\infty}^{+\infty} d \omega   \boldsymbol{\widetilde{\zeta}}(\omega) 2 \boldsymbol{\widetilde{\chi}}(\omega) \boldsymbol{\widetilde{\chi}}(-\omega)^T \\ &\boldsymbol{\widetilde{\zeta}}^T(-\omega) e^{ i \omega (t_1-t_2) } \\
= & \frac{1}{2 \pi}  \int_{-\infty}^{+\infty} d \omega  2 \mathbf{I} e^{ i \omega (t_1-t_2) }  = 2  \delta(t_1-t_2) .
\label{supp:xi_corr2}
\end{split}
\end{align}
It follows that $\boldsymbol{\xi}(t)$ is a delta-correlated stochastic
process, thus generalizing the work by Kawai~\cite{Kawai2015} to a tensorial
memory kernel.

As we adopted an approximation of $\boldsymbol{\theta } (t)$ whose components
are in the exponential form $\theta_{i,j} (t) = \sum_k A_{k,ij} e^{B_{k,ij}
	(t)} $, its Fourier transform is given by:
\begin{align}
\boldsymbol{\widetilde{\theta} } (\omega) = \sum_{k} \left[ -i \mathbf{A}_k \oslash \left( \omega \mathbf{J} + i \mathbf{B}_k \right) + i \mathbf{A}_k \oslash \left( \omega \mathbf{J} - i \mathbf{B}_k \right) \right] ,
\label{supp:eq:theta}
\end{align}
where $\oslash$ indicates the Hadamard division and $\mathbf{J}$ is an
all-ones $n \times n$ matrix.
Since $\boldsymbol{\theta}(t)$ is a real and even function of $t$,
$\boldsymbol{\widetilde{\theta} } (\omega)$ has to be real and even for real
values of $\omega$.
As a consequence, the singular points of $\boldsymbol{\widetilde{\theta} }
(\omega)$ have to be symmetric with respect to the real and imaginary axes,
namely in the form of pairs, $\pm i \mathbf{B}_k$.
For the same reason, the roots of $\boldsymbol{\widetilde{\theta} } (\omega)$ have to be symmetric with respect to the real and imaginary axes.
Thus, putting Eq.~\eqref{supp:eq:theta} into a common denominator,
factorizing, and using $\boldsymbol{\beta}_n$ and $\boldsymbol{\beta}_n^*$ to
denote the conjugate matrices containing the zeros of the numerator, yields:
\begin{align}
\begin{split}
\boldsymbol{\widetilde{\theta} } (\omega) = & \mathbf{K} \odot  \left(  \prod_n \left( \omega \mathbf{J}  -  \boldsymbol{\beta}_n \right) \odot  \left(\omega \mathbf{J}  - \boldsymbol{\beta}_n^* \right) \right) \\ &\oslash
\left( \prod_k \left( \omega \mathbf{J} + i \mathbf{B}_k \right) \odot \left( \omega \mathbf{J} - i \mathbf{B}_k \right) \right) ,
\label{supp:eq:theta1}
\end{split}
\end{align}
where $\odot$ is the Hadamard product, $\mathbf{K}$ is a matrix of positive
real numbers and it is assumed that ${\rm Im}(\beta_n)>0$ and ${\rm
	Im}(\beta_n^*)<0$.
It is worth noticing that since $\boldsymbol{\widetilde{\theta} } (\omega)$
is non-negative $\mathbf{K}$ contains positive values only~\cite{Kawai2015}.	
Define now the function $\boldsymbol{\widetilde{\chi}}(\omega)$ as:
\begin{align}
\begin{split}
\boldsymbol{\widetilde{\chi}}(\omega) =  &\frac{\mathbf{K}^{1/2}}{\sqrt{2}} \odot \langle \boldsymbol{ \mathcal{O} } , \boldsymbol{ \mathcal{O} } \rangle^{\odot 1/2}  \odot \prod_n i \left(\omega \mathbf{J} - \boldsymbol{\beta}_n^* \right) \\
& \oslash  \prod_k i \left(\omega \mathbf{J} + i \mathbf{B}_k\right) = \\
&\sum_{k} -i   \left(\omega \mathbf{I} + i \mathbf{B}_k\right)^{-1}  \mathbf{b}_k .
\end{split}
\label{supp:eq:g1}
\end{align}
Equation~(\ref{supp:eq:g1}) has to be solved to find the matrices
$\mathbf{b}_k$.

In the case of diagonal memory kernel matrix, $\mathbf{b}_k$ can be
straightforwardly obtained by solving:
\begin{align}
\begin{split}
\frac{\mathbf{K}^{1/2}}{\sqrt{2}} \langle \boldsymbol{ \mathcal{O} } , \boldsymbol{ \mathcal{O} } \rangle^{1/2} & \prod_n i \left(\omega \mathbf{I} - \boldsymbol{\beta}_n^* \right)   \prod_k i \left(\omega \mathbf{I} + i \mathbf{B}_k\right)^{-1} = \\
& \sum_{k} -i   \left(\omega \mathbf{I} + i \mathbf{B}_k\right)^{-1}  \mathbf{b}_k .
\end{split}
\label{supp:eq:g2}
\end{align}

Finally, for a one-dimensional GLE the presented formulation reduces to the
one derived by Kawai~\cite{Kawai2015}, therefore the coefficients $b_k$ can
be evaluated from the following relation:
\begin{align}
\langle \boldsymbol{ \mathcal{O} } , \boldsymbol{ \mathcal{O} } \rangle A_k=- 2 b_k \sum_{n} \frac{b_n}{B_k +B_n} ,
\end{align}
obtained by Eqs~\eqref{supp:eq:g}, \eqref{supp:eq:theta-g} and \eqref{supp:eq:theta}.

\paragraph{Extended dynamics and integration algorithm}

For a general $\boldsymbol{\theta}(t)$, we have the following extended
dynamics:
\begin{align}
\begin{cases}
\partial_t \boldsymbol{ \mathcal{O} }(t) = \mathcal{P} \mathcal{L} \boldsymbol{ \mathcal{O} } - \int_{0}^{t}  \boldsymbol{\theta}(\tau) \boldsymbol{ \mathcal{O} } (t-\tau) d \tau + \sum_k \mathbf{R}_{k}(t) , \\
\partial_t \mathbf{R}_k(t)=  \mathbf{B'}_k  \mathbf{R}_k(t) + \mathbf{b}_k \boldsymbol{\xi}(t) ,
\end{cases}
\end{align}
where the convolution can be decomposed in different ways depending on the structure of $\boldsymbol{\theta}(t)$.
In our case $\boldsymbol{\theta}(t)$ has an exponential form, thus
$\mathbf{B'}_k=\mathbf{B}_k$ and the variables $\mathbf{S}_k(t)=
-\mathbf{Z}_k(t) + \mathbf{R}_k(t)$ can be defined so that the GLE is
rewritten in the following form:
\begin{align}
\begin{cases}
&\partial_t \boldsymbol{ \mathcal{O} }(t) = \mathbf{F} (\boldsymbol{ \mathcal{O} }(t)) + \sum_{k=1}^{N_n} \mathbf{S}_{k}(t) , \\
&\partial_t \mathbf{S}_{k}(t)= \mathbf{B}_k \mathbf{S}_{k}(t) - \mathbf{A}_k  \boldsymbol{ \mathcal{O} } (t) + \mathbf{b}_k  \boldsymbol{\xi}(t),
\end{cases}
\label{system-to-be-solved}
\end{align}
with $\mathbf{F}(\boldsymbol{ \mathcal{O} }(t)) =  \mathcal{P} \mathcal{L} \boldsymbol{ \mathcal{O} } $ accounting for the conservative mean force contributions.

The numerical algorithm adopted to solve the system in
Eq.~\eqref{system-to-be-solved} involves a splitting scheme together with the
Euler-Maruyama scheme for the stochastic part, $S_k(t)$:
\begin{align}
&\boldsymbol{ \mathcal{O} }^{\left(n+1/2\right)}=\boldsymbol{ \mathcal{O} }^{\left(n\right)}+ \frac{\Delta t}{2} F^{c}(\boldsymbol{ \mathcal{O} }^{\left(n\right)})  +  \frac{\Delta t}{2} \sum_{k=1}^{N_n} S_{k}^{\left(n\right)} , \\
&S_{k}^{\left(n+1\right)}= \left( 1+B_k \Delta t \right) S_{k}^{\left(n\right)} - A_k \boldsymbol{ \mathcal{O} }^{\left(n+1/2\right)} \Delta t +  b_k \xi_{k}^{\left(n\right)}  ,\\
&\boldsymbol{ \mathcal{O} }^{\left(n+1\right)}=\boldsymbol{ \mathcal{O} }^{\left(n+1/2\right)}+ \frac{\Delta t}{2} F^{c}(\boldsymbol{ \mathcal{O} }^{\left(n+1\right)})  +  \frac{\Delta t}{2} \sum_{k=1}^{N_n} S_{k}^{\left(n+1\right)} ,
\end{align}
where ${\xi}_{k}^{\left(n\right)}\sim \mathcal{N} (0, 2 \Delta t)$ are independent Gaussian distributed random values.

To test the numerical stochastic integrator, similarly to
Ref.~\cite{Baczewski2013}, we consider a one-dimensional GLE with a
single exponential memory kernel and no conservative forces.
In this specific case, the time correlation is analytically solvable:
\begin{align}
\frac{\langle \boldsymbol{ \mathcal{O} }(t) \boldsymbol{ \mathcal{O} }(0) \rangle}{\langle \boldsymbol{ \mathcal{O} }(0) \boldsymbol{ \mathcal{O} }(0) \rangle}=
\begin{cases}
e^{ \frac{tB}{2}} \left( \cos(\Omega t) - \frac{B}{2 \Omega}\sin(\Omega t) \right) \quad & \Omega \neq 0 , \\
e^{ \frac{tB}{2}} \left( 1- \frac{B t}{2} \right) \quad &\Omega = 0 ,
\end{cases}
\end{align}
where the complex parameter $\Omega= \sqrt{A- B^2 /4}$ was introduced.
Figure~\ref{fig:autv} shows that the numerical integrator is able to
accurately reproduce the analytical correlation in the under-damped limit
($A=1$ and $B=1$), in the damped case ($A=1$ and $B=-2$) and in the
over-damped limit ($A=1$ and $B=-4$).

\begin{table}[t]
	\centering
	\caption{Values of the interaction potentials parameters, adopted to simulate the particle chain.}
	\begin{tabular}{| c | c  c c c c c c  |}
		\hline
		Parameters & $K_H$ & $K_\gamma$ & $K_\phi$ & $r_0$ & $\gamma_0$ & $\epsilon_{i,j}$ & $\sigma_{i,j}$ \\ \hline
		Values & 100 & 10 & 10 & 1.5 & 109.5 &1 &1  \\ \hline
	\end{tabular}
	\label{table:chain}
\end{table}

\section{Single particle in a bath: simulation details and additional results}

\begin{figure}[h]
	\centering
	\centering
	\includegraphics[width=0.3\textwidth]{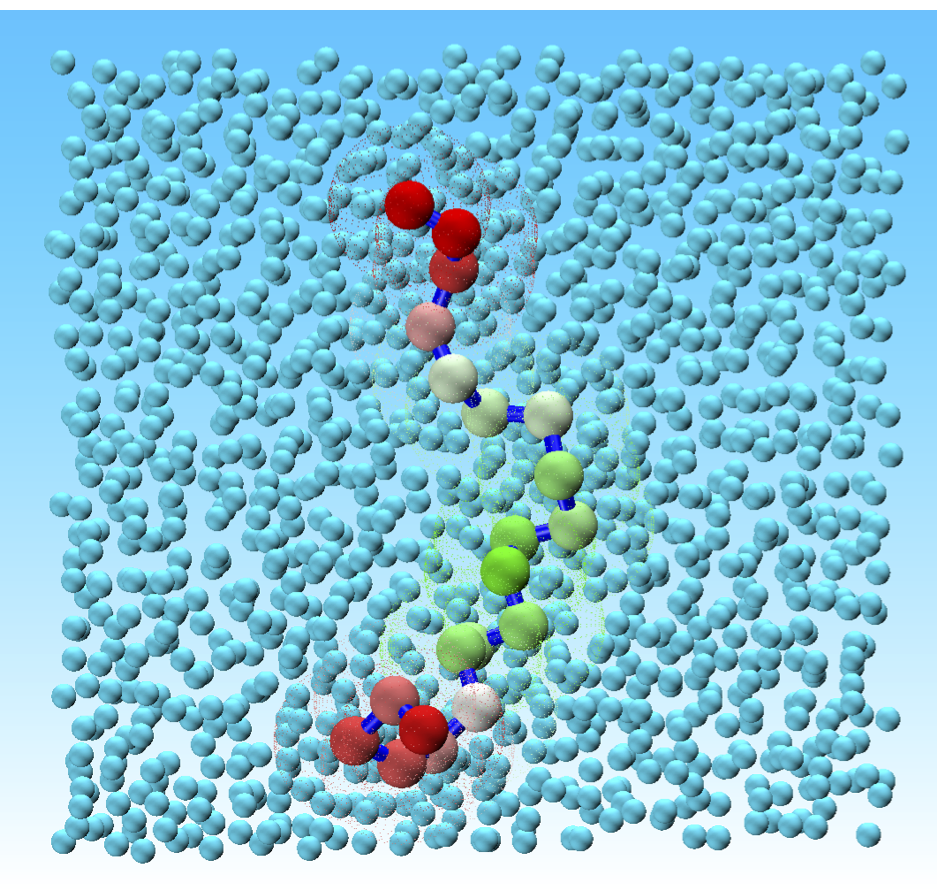}
	\caption{Representation of a particle chain at equilibrium with a heat bath.}
	\label{fig:chain_image}
\end{figure}

A target colloidal particle, with mass $m=1$, is immersed in a bath of
identical particles with masses $m_b=1$.
Two systems are studied. We simulate a low density limit (LDL) with $700$
particles in total, while the high density limit (HDL) with $800$ particles.
The interaction between any two particles $i$ and $j$ is modelled by the
Lennard-Jones (LJ)potential.
%
MD simulations were performed integrating particles governing equations in time by using a Verlet algorithm.
The time step is fixed at $\Delta t= 10^{-3}$.
The following procedure is then adopted for the MD simulations.
First, the bath particles are randomly generated inside the simulation box.
Then, a minimization algorithm is employed to avoid overlaps between particles.
Hence, a run of $10^{5}$ time steps is used to equilibrate the system.
Finally, data on forces and momenta are gathered over $10^{5}$ time steps.
This process is repeated for $10^{2}$ trajectories in order to enhance the
accuracy of the correlations, and consequently, of the memory kernels.

In Figs~\ref{fig:bath_theta1}(a,b) we report the mean-square displacement (MSD), $\langle \left( x(t) -
x(0) \right)^2 \rangle$, computed with MD, LE and GLE in the LDL and HDL
cases.
It is evident that both Markovian and non-Markovian coarse grainings are able
to accurately reproduce the MSD.
Moreover, in the HDL case, the GLE shows better performance with respect to
the LE.
Figures~\ref{fig:bath_theta1}(c,d) depict the values of the adaptive
learning rate $\eta$ during the MLP learning process.
The log-log plot highlights the wide range of $\eta$ values spanning up to
$8$ orders of magnitude.
This variability exemplifies the advantages of an adaptive learning rate over a
fixed one.
The cost function $C$ evolution during the learning process is reported in
Figs~\ref{fig:bath_theta1}(e-f).
The monotonically decreasing trend of $C$ shows a plateau at some point which
corresponds to the end of the learning process.

We now turn our attention to the performance of the adopted coarse-graining
out of equilibrium by analyzing the PDF, $\rho$. A target particle with zero
initial position $x$ and momentum $p$ is immersed in an equilibrated bath of
$699$ particles identical to the one adopted in the LDL case at equilibrium.
$10^5$ trajectories relaxing to equilibrium are simulated.
This relaxation corresponds to the evolution of a Dirac delta to the
equilibrium distribution in the phase space.
Similarly, the relaxation of $\rho$ obtained by coarse-graining the bath with
both GLE and LE is followed.
The comparison reported in Figs~\ref{fig:bath_nneq_pdf}(a-f) shows that GLE,
even if parameterized with a memory kernel evaluated at equilibrium
conditions, significantly outperforms LE.
As expected, at equilibrium the distributions obtained with MD, GLE and LE converge.
During the relaxation, $\rho$ relaxes faster for LE and GLE with respect to MD.
A quantitative estimation of the accuracy of GLE in reproducing the density
relaxation is provided by the mean square errors in position $\epsilon_q$ and
momentum $\epsilon_p$, shown in Figs~\ref{fig:bath_nneq_pdf}(g,h).
As expected, both errors are negligible at the start and asymptotically for
large times when the system reaches equilibrium.
During the initial stage of the relaxation, the errors $\epsilon_q$ and
$\epsilon_p$ approach a peak, whose value for GLE is lower than that for LE
by about $50\%$ and $35\%$, respectively.

\section{Particle chain in a bath: simulation details}

A chain of $N=20$ colloidal particles in an LJ bath is also simulated, as
shown in Fig.~\ref{fig:chain_image}.
%
%
%
For the multi-body Dreiding potential~\cite{Mayo1990,Hossain2010}, linear covalent bonds are approximated by the harmonic potential
$v_{H}(\mathbf{r}_{ij})=k_H (\mathbf{r}_{ij} - \mathbf{r}_0)^2$, where
$\mathbf{r}_0$ is the equilibrium position and $k_H$ is a positive constant.
Similarly, angular covalent bonds are approximated by $v_{\gamma}(r_{ijk})=
k_\gamma (\gamma_{ijk} - \gamma_0)^2$, where $\gamma_{ijk}$ is the angle in
$i$ formed by the particles $i$, $j$ and $k$, $\gamma_0$ is the equilibrium
angle and $k_\gamma$ is a positive constant.
Finally, we torsional (dihedral) bonds are modelled through the potential $v_{\phi}(r_{ijkl})= k_\phi (1+ cos(2\phi_{ijkl}))$, with $\phi_{ijkl}$ being the angle between the two planes defined by $\left \{ \mathbf{r}_i, \mathbf{r}_j, \mathbf{r}_k \right \} $ and $\left\{ \mathbf{r}_j, \mathbf{r}_k, \mathbf{r}_l \right\} $ respectively, and $k_\phi$ being a positive parameter.
Table~\ref{table:chain} gives the values of all intermolecular parameters.
The bath contains $69,900$ particles interacting with Lennard-Jones potential
$v_{LJ}$. The simulation box measures $50 \times 50 \times 40$ in reduced
units and periodic boundary conditions are imposed along the $x$, $y$ and $z$
axes.
A Nos\'e-Hoover thermostat is used to equilibrate the system at a reduced temperature $T=1.1$ with a time step $\Delta t= 10^{-2}$.

The following procedure is followed to run the MD simulations.
First, the bath particles are randomly generated inside the simulation box.
Then, the chain particles are placed along a straight line and a minimization
algorithm is employed to avoid overlaps between them.
Accordingly, a run of $1.5 \times 10^{5}$ time steps is used to equilibrate
the system. Finally, data are gathered over $2 \times 10^{7}$ time steps.

\section*{GLE for time series}

To model a general time series of an observable $\boldsymbol{ \mathcal{O} }$
by means of a non-Markovian GLE the following conditions have to be
satisfied:
\begin{itemize}
	\item $\boldsymbol{ \mathcal{O} }(t) \sim \mathcal{N}(\mu,\,\sigma^{2})$ ,
	\item $ \langle \boldsymbol{ \mathcal{O} }(t) \rangle = 0 \quad \forall  t  $ ,
	\item $\langle  \boldsymbol{ \mathcal{O} }^2 (t) \rangle = \sigma^{2}  \quad \forall  t $ ,
	\item $ \langle \boldsymbol{ \mathcal{O} }(t) \boldsymbol{ \mathcal{O} }(t') \rangle = \langle \boldsymbol{ \mathcal{O} }(t-t') \boldsymbol{ \mathcal{O} }(0) \rangle \quad \forall  t \geq t' $ .
\end{itemize}
If the original data of $\boldsymbol{ \mathcal{O} }$ exhibits non-stationary
features, some manipulation of the data is needed to obtain stationarity.

\paragraph*{Case 1: Global temperature dynamics}

\begin{table}[t]
	\centering
	\caption{Results of the augmented ADF test for modified global temperature.}
	\begin{tabular}{| c | c |}
		\hline
		ADF Statistic:  &  $-21.377945$ \\ \hline
		p-value: 			& $<10^{-16}$  \\ \hline
		lags: 			& $54$  \\ \hline
		Critical Values: & \\
		1\%: & $-2.566$ \\
		5\%: & $-1.941$ \\
		10\%: & $-1.617$ \\ \hline
	\end{tabular}
	\label{table:ADF_temp}
\end{table}

\begin{figure}[h]
	\centering
	\begin{minipage}{0.4\textwidth}
		\flushleft (a)\\
		\centering
		\includegraphics[width=\textwidth]{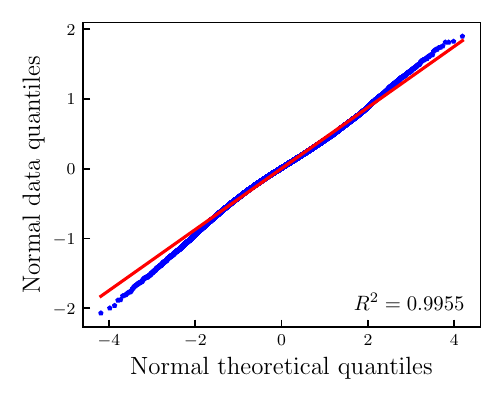}
	\end{minipage}
	\begin{minipage}{0.4\textwidth}
		\flushleft (b)\\
		\centering
		\includegraphics[width=\textwidth]{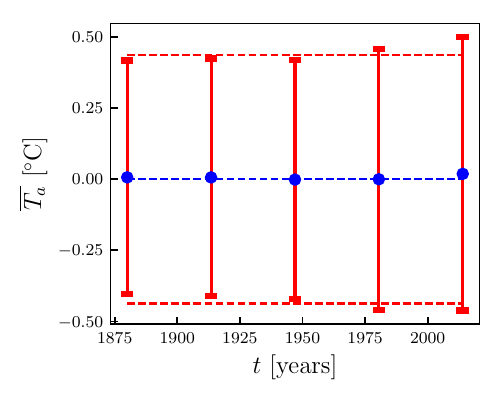}
	\end{minipage}
	\begin{minipage}{0.4\textwidth}
		\flushleft (c)\\
		\centering
		\includegraphics[width=\textwidth]{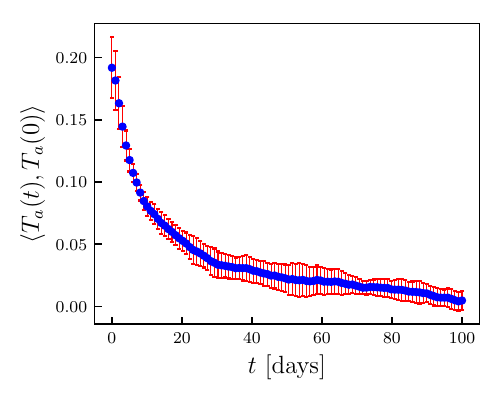}
	\end{minipage}
	\caption{(a) QQ-plot for $T_a(t)=T(t)-T_y(t)$. The quantiles obtained from the distribution of the data are plotted against the
		quantiles computed directly from the normal distribution.
		(b) Mean of $T_a$ evaluated for 5 different data windows in time (blue dots) and corresponding standard deviations represented as red error bars.
		(c) Average time correlation function (blue dots) and standard error evaluated at each time from the time correlations of 5 different data windows.
	}
	\label{fig:supp:temp}
\end{figure}


The yearly moving average is defined as:
\begin{align}
& T_y (t) = \frac{1}{y}\sum_{i=t-y-1}^{t-1} T (i) .
\label{eq:yearly_moving}
\end{align}
%

To test the statistical properties ot $T_a$, we adopted the following tests.
First, we employ the Q–Q (quantile-quantile) plot which compares two
distributions by plotting their quantiles against each other.
Quantiles are defined as sets of values of a stochastic variable splitting a
distribution into an arbitrary number of intervals with identical
probability.
Figure~\ref{fig:supp:temp}(a) shows the QQ (quantile-quantile) plot which
compares the data cumulative distribution against the normal theoretical
cumulative distribution for each quantile.
The red straight line represents the case in which the distribution of the
data is exactly normal.
Evidently, the time series data are well approximated by a normal
distribution, especially in the theoretical quantile range $-3<Q<3$. Some
tail effects are visible, but the overall agreement is quantitatively
verified by the $R$-squared test which gives a value $R^2= 0.9955$.

To test now the stationarity of mean variance, and time correlations, we
split the data in 5 windows.
Figure~\ref{fig:supp:temp}(b) shows that, assuming the observable of interest
$T_a(t)$ is stationary, the maximum errors for the mean and standard
deviation are $0.0183$ and $0.0430$, respectively.
Moreover, as reported in Fig.~\ref{fig:supp:temp}(c), the maximum standard
error between the time correlation for each window and their mean is
$0.0246$.

Finally, to test the stationarity of the modified time series, the augmented
Dickey–Fuller (ADF) test is adopted. 
The est is useful to establish if a unit root is present in the stochastic
data series. Specifically, the null hypothesis of a unit root is rejected in
favor of the stationary alternative if the test statistic is more negative
than some critical values.
The results of ADF reported in Table~\ref{table:ADF_temp} allows us to reject
the unit root hypothesis with a probability higher than $99\%$.

\paragraph*{Case 2: Nikkei index}

\begin{table}[h]
	\centering
	\caption{Results of the augmented ADF test for modified Nikkei index.}
	\begin{tabular}{| c | c |}
		\hline
		ADF Statistic:  &  $-29.805726$ \\ \hline
		p-value: 			& $<10^{-16}$  \\ \hline
		lags: 			& $10$  \\ \hline
		Critical Values: & \\
		1\%: & $-2.566$ \\
		5\%: & $-1.941$ \\
		10\%: & $-1.617$ \\ \hline
	\end{tabular}
	\label{table:ADF_nikkei}
\end{table}

The moving average and the moving standard deviation
are computed over a period $[t-y,t-1]$ as:
\begin{align}
& NI_y (t) = \frac{1}{y} \sum_{i=t-y-1}^{t-1} NI (i) , \\
& \sigma_y (t) = \sqrt{ \frac{1} {y} \sum_{i=t-y-1}^{t-1} \left( NI (i) - NI_y (t) \right)^2 } .
\label{eq:moving}
\end{align}
The parameter $y$ is appropriately chosen in order to obtain a stationary
$NI_a (t)$; preliminary tests have shown that $y=10$ days is an optimal value.
%

\begin{figure}[t]
	\centering
	\begin{minipage}{0.4\textwidth}
		\flushleft (a)\\
		\centering
		\includegraphics[width=\textwidth]{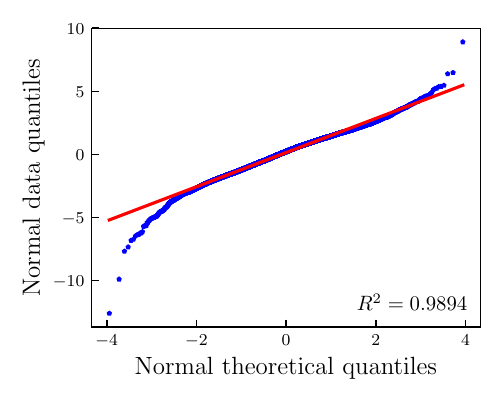}
	\end{minipage}
	\begin{minipage}{0.4\textwidth}
		\flushleft (b)\\
		\centering
		\includegraphics[width=\textwidth]{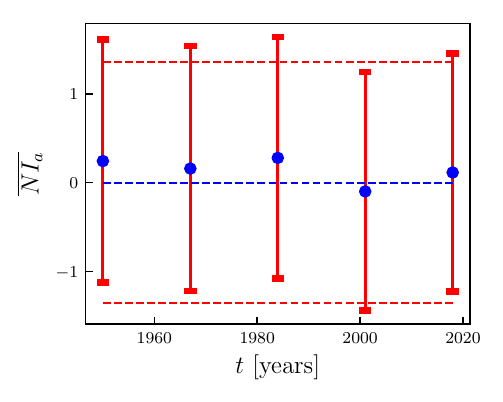}
	\end{minipage}
	\begin{minipage}{0.4\textwidth}
		\flushleft (c)\\
		\centering
		\includegraphics[width=\textwidth]{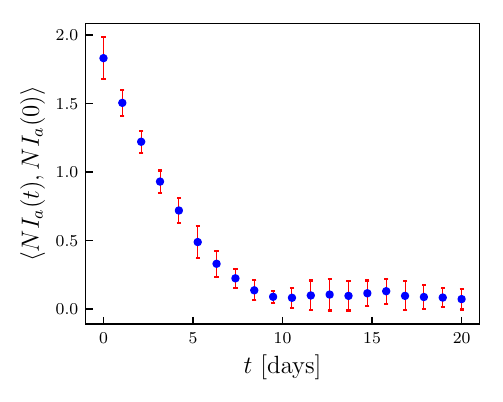}
	\end{minipage}
	\caption{(a) QQ-plot for $NI_a(t)=NI(t)-NI_y(t)$.
		(b) Mean of $NI_a$ evaluated for 5 different data windows in time (blue dots) and corresponding standard deviations represented as red error bars.
		(c) Average time correlation function (blue dots) and standard error evaluated at each time from the time correlations of 5 different data windows.
	}
	\label{fig:supp:nikkei}
\end{figure}

Figure~\ref{fig:supp:nikkei}(a) shows the QQ plot.
The time series distribution is well approximated with a normal distribution
in the theoretical quantile range $-2.5<Q<2.5$, but evidently, strong tail
effects are present.
This means that the Gaussian approximation, and consequently the GLE for
$NI(t)$, remains valid as long as extreme market events, such as market
crashes or crises, are avoided.
The overall agreement is quantitatively verified by the $R$-squared test,
which gives $R^2= 0.9894$.

Finally, to test the stationarity of mean variance, and time correlations, we
split the data in 5 equally sized sets and for each one we analyze their
statistical properties.
Figure~\ref{fig:supp:nikkei}(b) shows that, taking $NI_a$ as the observable,
the maximum errors for the mean and standard deviation are $ 0.2787 $ and
$0.0234 $, respectively. But unlike the global temperature problem we
considered earlier, the error for the mean is an order-of-magnitude higher,
suggesting that stock markets are more challenging to model than natural
phenomena. Moreover, as reported in Fig.~\ref{fig:supp:nikkei}(c), the
maximum standard error between the time correlation in each window and their
mean is $ 0.1082 $.
The results of the ADF test reported in Table~\ref{table:ADF_nikkei} allow us
to reject the unit root hypothesis with a probability higher than $99\%$.

\section*{Acknowledgment}
We acknowledge financial support from the Imperial College Chemical Engineering PhD Scholarship scheme, European Research Council via Advanced Grant No. 247031, Engineering and Physical Sciences Research Council of the UK via grants No. EP/L025159 and EP/L020564 and the Defense Advanced Research Projects Agency of the USA.

\bibliographystyle{plain}
\bibliography{Refs}

\end{document}